\documentclass[bezier,11pt]{article}

\usepackage{amsfonts}
\usepackage{amsmath, amssymb, latexsym}
\usepackage{mathrsfs}
\usepackage{rotating}
\usepackage{amsthm} 
\usepackage{bbm}
\usepackage{textcomp}
\usepackage{mathabx}

\newtheorem{theorem}{Theorem}[section]
\newtheorem{corollary}[theorem]{Corollary}
\newtheorem{assumption}[theorem]{Hypothesis}

\newtheorem{remark}[theorem]{Remark}
\newtheorem{definition}[theorem]{Definition}

\newtheorem{lemma}[theorem]{Lemma}
\newtheorem{proposition}[theorem]{Proposition}

\newcommand {\bd}{\begin{definition}}
\newcommand {\ed}{\end{definition}}
\newcommand {\bpro}{\begin{proposition}}
\newcommand {\epro}{\end{proposition}}
\newcommand {\bl}{\begin{lemma}}
\newcommand {\el}{\end{lemma}}
\newcommand {\bcor}{\begin{corollary}}
\newcommand {\ecor}{\end{corollary}}
\newcommand {\brem }{\begin{remark} \rm }
\newcommand {\erem }{\end{remark}}
\newcommand{\bethe}{\begin{theorem}}
\newcommand{\ethe}{\end{theorem}}
\newcommand {\bassumption}{\begin{assumption}}
\newcommand {\eassumption}{\end{assumption}}

\numberwithin{equation}{section}

\newcommand{\dcb}{\begin{array}{lll}}
\newcommand{\dce}{\end{array}}
\newcommand{\ebe}{\begin{enumerate}\setlength{\baselineskip}{14pt}\setlength{\parskip}{0pt}}
\newcommand{\dbe}{\end{enumerate}}
\def \ind{1\!\!1\!}
\def\cro#1{\langle #1\rangle}

\newcommand{\guillemetdb}{\texttt{\textquotedblright}}

\def\twoprimes#1{\texttt{\textquotedblright}#1\texttt{\textquotedblright}}

\newcommand{\replacegamma}{{^{\circ}}\!Z}
\newcommand{\replacemdot}{\mathsf{m}}
\newcommand{\replacemathsfm}{\mathsf{s}}
\newcommand{\replacem}{\mathsf{M}}
\newcommand{\replaceMtildec}{\mathsf{r}}
\newcommand{\replaceHX}{{^R}\!|\!}
\newcommand{\replaceddotxi}{\overline{\xi}}
\newcommand{\replacedotzeta}{{^{\circ}}\!\zeta}
\newcommand{\replacemathsfd}{\mathsf{b}}

\newcommand{\lc}{[\![}
\newcommand{\rc}{]\!]}

\begin{document}

\

\begin{center}
\textbf{\Large Local martingale deflators for asset processes stopped at a default time $S^{\tau}$ or right before $S^{{\tau}-}$}

Shiqi Song\footnote{This research has benefited from the support of the ``Chair Markets in Transition'',
F\'ed\'eration Bancaire Fran\c caise, and of the ANR project 11-LABX-0019.}\\
Laboratoire Analyse et Probabilit\'{e}s\\ 
Universit\'{e} d'Evry Val D'Essonne, France
\end{center}

\

\begin{abstract}
Since the seminal paper \cite{DSS}, people knows that, to use the hazard rate to evaluate the price $S$ of securities defaultable at $\tau$, one should firstly find the solution $V$ of a specific backward stochastic differential equation written with the hazard rate process, and then apply the pre-default formula $S\ind_{[0,\tau)}=V\ind_{[0,\tau)}$, whenever $\Delta_\tau V=0$. This pricing formula has prompted a lot of discussions, especially on the condition $\Delta_\tau V=0$.  The method by backward stochastic differential equation has been adopted in recent works \cite{CS1,CS2} to evaluate the counterparty risk and funding cost. The jump problem $\Delta_\tau V=0$, however, has been avoided, because a filtration reduction and a probability change technique under the name \twoprimes{condition (A)} has been introduced in \cite{CS1,CS2}, which allow one to solve directly the backward stochastic differential equation satisfied by $S^{\tau-}$. 

Our future aim is to extend the technique of \cite{CS1,CS2} to cover more general models. A first step towards this extension is to generalize condition (A), when probability measure changes are replaced by the local martingale deflator changes. Concretely in this paper, we consider a pair of filtrations $\mathbb{F}\subset \mathbb{G}$ which become different {only from the default time $\tau$ onwards}. For an $\mathbb{F}$ semimartingale $S$ having an $\mathbb{F}$-deflator, we establish conditions on $S$ such that $S^{\tau-}$ can have a $\mathbb{G}$-deflator. Under these conditions, we construct $\mathbb{G}$-deflators for $S^{\tau-}$ in term of Az\'ema supermartingale of $\tau$. (In passing, the same for $S^\tau$ will also be done.)

Our study is based on the existence of a subfiltration $\mathbb{F}$ which \twoprimes{coincides} with $\mathbb{G}$ on $[0,\tau)$. For applications, it is important to have a method to infer the existence of such a filtration $\mathbb{F}$ from the knowledge of the market information $\mathbb{G}$ and from the default time ${\tau}$. This question is discussed at the end of the paper.

\

\textbf{Key words:}
no-arbitrage of the first kind, local martingale deflator, progressive enlargement of filtration, credit risk modeling, counterparty risk.

\

\textbf{MSC 2010 numbers:} 60G07, 60G44, 91G40.

\

\textbf{JEL classification code:} G11, G14, G32.

\

$^*${\footnotesize version du 30 novembre 2014}
\end{abstract}

\pagebreak

\section{Introduction}

\subsection{Results}

Consider a credit risk model $(\mathbb{G},\tau, S)$, where $\mathbb{G}$ is a filtration, $\tau$ is a $\mathbb{G}$ stopping time, and $S$ is a multi-dimensional process. Let $\mathbb{F}$ be a subfiltration of $\mathbb{G}$ satisfying the two conditions: 
\ebe\item[.]
$\mathbb{F}$ \texttt{"}coincides\texttt{"} with $\mathbb{G}$ up to the time $\tau$ (cf. Assumption \ref{reduction}); 
\item[.]
$S$ is an $\mathbb{F}$ semimartingale possessing an $\mathbb{F}$ local martingale deflator (cf. Definition \ref{defdef} below). 
\dbe
(Note that the first condition is introduced in \cite{CS1} and constitutes a generalization of the usual setting of progressive enlargement of $\mathbb{F}$ with the random time $\tau$.) In this paper we study the problem :
\begin{quote}\itshape
\textbf{Deflator problem.} Under what condition, the process $S^{\tau-}$ (stopped right prior to $\tau$) or $S^\tau$ possesses a local martingale deflator in $\mathbb{G}$. 
\end{quote}
We will set up an appropriate theoretical framework, within which the above question finds answers. Before presenting the motivations in the next subsection, here are the principal results.
\ebe
\item[$_\bullet$]
A necessary and sufficient condition is given in Theorem \ref{whenS} for $S^{\tau-}$ to possess a local martingale deflator in $\mathbb{G}$. This result is obtained only after various preliminary results have been proved, as indicated below.

\item[$_\bullet$]
The deflator problem depends, in a delicate way, on the Az\'ema supermartingale $Z$ of the random time $\tau$ in $\mathbb{F}$. Two decompositions of $Z$ are involved in our work: the Doob-Meyer's decomposition $Z={\replacemdot}+\mathtt{a}$ and the predictable multiplicative decomposition $Z=LD$ defined in \cite{jacod}, with a multiplicative martingale part $L$ and a multiplicative drift part $D$. This multiplicative decomposition is valid only on the set $\{{^{p\cdot\mathbb{F}}}\!(Z)>0\}$ ($\mathbb{F}$ predictable projection of $Z$). For the construction of deflators, we need to extend the decomposition $Z=LD$ onto the whole $\mathbb{R}_+$ and ensure that $d\mathsf{a}_s$ does not charge the set $\{L_-=0\}$. Also, the first zero time $\zeta$ of $Z$ is to be decomposed into three elements $\eta,\dot{\eta}, \ddot{\eta}$, each of which concerns a different part of deflators of $S^{{\tau}-}$. We have to control which components of $Z$ can vanish at these stopping times. In section \ref{ZP}, a full investigation is made on the processes $L,D, Z_-, {^{p\cdot\mathbb{F}}}\!(Z)$ in association with the stopping times $\eta,\dot{\eta}, \ddot{\eta}$. As a consequence of this study, we establish in Lemma \ref{newMdecomp} the formula (different from the one in \cite{jacod}) $D=Z_0\mathcal{E}(-\frac{1}{Z_-}{_\centerdot}\mathsf{a})$.  

\hspace{17pt}This formula of $D$, jointly with the reduction results of section \ref{Freduction} (especially the equation of Lemma \ref{yyam}) forms a passage connecting the multiplicative decomposition $Z=LD$ to the deflator computation of $S^{\tau-}$. Recall that one of the theoretical conjectures about the deflator problem for $S^{{\tau}-}$ is that $\mathbb{G}$-deflator is a multiple of $\mathbb{F}$-deflator by a factor inversely proportional to $L$ (the predictable multiplicative martingale part of $Z$). With the formula of $D$ and the reduction results, this conjecture is confirmed in subsection \ref{generalsm} by that there exists a factor $M$ depending on $S$ such that $\mathbb{G}$-deflator of $S^{{\tau}-}$ is the multiplication of $\mathbb{F}$-deflator with $\frac{M^{{\tau}-}}{L^{{\tau}-}}$. It is interesting to compare this multiplicative property with the result of \cite{AFK} that proves that the $\mathbb{G}$-deflator of $S^{{\tau}}$ (instead of $S^{{\tau}-}$) is inversely proportional to the \underline{optional} (instead of \underline{predictable}) multiplicative martingale part of $Z$ (cf. \cite{K2014}). 

\hspace{17pt}The study in section \ref{ZP} enriches a lot our knowledges on the Az\'ema supermartingale and is useful in general. See \cite{CS1} for other applications of the formula of $D$.

\item[$_\bullet$]
The well-known formula of Jeulin-Yor (cf. \cite{DM3,Jeulin80,JY}) gives the $\mathbb{G}$ semimartingale decomposition of $X^\tau$ for $\mathbb{F}$ local martingale $X$. However, the study of this paper requires results which characterize the $\mathbb{G}$ local martingales with their reductions in $\mathbb{F}$. Section \ref{Freduction} is devoted to the reduction problem with formulas in Lemma \ref{rdm}, Lemma \ref{yyam}, Lemma \ref{csinv}, which give the above mentioned characterization of the $\mathbb{G}$ local martingales \guillemetdb living on $[0,\tau)$\texttt{"}. The reduction result is the cornerstone of subsection \ref{generalsm} (which implies Theorem \ref{whenS}). It is also an essential element in \cite{CS1}.

\item[$_\bullet$]
In the article \cite{AFK}, the local martingale deflators for $S^\tau$ (instead of $S^{\tau-}$) in $\mathbb{G}$ are considered. The proof of the main result Theorem 1.2 in \cite{AFK} raises the following question : supposing that an $\mathbb{F}$ semimartingale $S$, without jump at a given stopping time $R$ ($\Delta_RS=0$), possesses a $\mathbb{F}$ local martingale deflator $Y$, can one construct a (second) local martingale deflator for $S$ which has no jump at $R$ neither ? This question, which remains open in \cite{AFK}, is answered in Theorem \ref{ddotxi}. The proof of the theorem is based on a subtle analysis of the stochastic logarithmic $\xi$ of $Y$, which improves our knowledge about deflators. Par example, it is proved that the drift of $\xi^{R-}$ (as $\xi$ itself) can not have too negative jumps.

\item[$_\bullet$]
Also for the proof of Theorem \ref{ddotxi}, we are led to study the $\mathbb{F}$ local martingales $X$ whose $X^{R-}$ remains local martingale. We prove in Lemma \ref{ortho}
that they are precisely the local martingales orthogonal to $\ind_{\{R>0\}}\ind_{[R,\infty)}-(\ind_{\{R>0\}}\ind_{[R,\infty)})^{p\cdot\mathbb{F}}$ ($\mathbb{F}$ compensated jump process). We then establish an orthogonal decomposition formula in Theorem \ref{Rdecomp}.

\item[$_\bullet$]
Theorem \ref{whenS} depends on Theorem \ref{ddotxi}, because, without it, the reduction results would not be applicable in subsection \ref{generalsm}. Another application of Theorem \ref{ddotxi} is that it enables us to work on \cite[Theorem 1.2]{AFK} under the original probability measure, without passing through the auxiliary probability change of \cite{AFK}. Therefore, a new proof of \cite[Theorem 1.2]{AFK} is given in Corollary \ref{newafk}. Moreover, we will apply the idea of \cite{songnote} to produce an explicit deflator for $S^\tau$ with the components of $Z$. It is a very different construction from \cite{AFK}, because it does not depend on the optional multiplicative decomposition of $Z$ established in \cite{K2014}.

\item[$_\bullet$]
The present work is based on the existence of a subfiltration $\mathbb{F}$ which coincides with $\mathbb{G}$ on $[0,\tau)$, but does not accept $\tau$ as stopping time. However, the existence of such a subfiltration is not unanimously	accepted, all the more so as no practical method exists to infer the possible presence of such a filtration $\mathbb{F}$. In section \ref{condB}, for the first time, some results will be proved in response to that controversial situation.

\dbe

\subsection{Motivations}

To apprehend our study of $S^{\tau-}$, we need to review quickly the pricing system of defaultable securities. According to arbitrage pricing principal, the price process of a defaultable security satisfies the formula$$
S_t=\mathbb{E}[\ind_{\{t<\tau\leq T\}}e^{-\int_t^\tau r_sds}Y_\tau+\ind_{\{T<\tau\}}e^{-\int_t^T r_sds}\xi|\mathcal{G}_t],
$$ 
under a neutral probability, with the short rate process $r$, a payoff $\xi$ and a recovery rule $Y$. This is however a unsatisfactory pricing formula, because it does not involve explicitly the hazard rate process $h$. In response, the paper \cite{DSS} provides the following formula (under few technical condition): $$
S_t=V_t,\ t<\tau,
$$ 
if $\Delta_\tau V=0$, where $V$ is the solution of a backward stochastic differential equation (BSDE, in abbreviated form):$$
V_t=\mathbb{E}[\int_t^Te^{-\int_t^s (r+h)_udu}Y_sh_sds+e^{-\int_t^T (r+h)_udu}\xi|\mathcal{G}_t].
$$
The advantage of $V$ over the initial formula of $S$ is explained in \cite{DSS} (cf. also \cite{CDGH}). Since then, much has been written on this subject, especially on the delicate jump condition $\Delta_\tau V=0$ (cf. \cite{BR}). The key observation is that, if we choose a suitable pre-default value process (i.e. a process that coincides with $S$ on $[0,\tau)$), it can be computed by a kind BSDE and the BSDE consideration is the best way to make appear the hazard rate process in the pricing formula.

This approach of pre-default value process with BSDE has been adopted in \cite{CS1,CS2} to evaluate the counterparty risk and funding cost $X$, with some important modifications. Firstly, \cite{CS1,CS2} has chosen the particular pre-default process $X^{\tau-}$ and established its BSDE$^X$. This BSDE$^X$ is not simple, because it requires its solution to satisfy the jump vanishing property $\Delta_\tau V=0$, which brings us back to the discussion of the formula in \cite{DSS}. A condition, called condition (A), is then introduced. 
\begin{quote}\itshape
(\textbf{A})\ There exists a subfiltration $\mathbb{F}\subset \mathbb{G}$, which coincides with $\mathbb{G}$ on $[0,\tau)$. There exists at the same time an equivalent probability measure $\mathbb{P}$ such that, for every $(\mathbb{F},\mathbb{P})$ local martingale $M$, $M^{\tau-}$ is a $(\mathbb{G},\mathbb{Q})$ local martingale. 
\end{quote}
It is then proved that, under condition (A) and the positivity of $Z$, there exists a classical $(\mathbb{F},\mathbb{P})$ BSDE$^\circ$ such that, for any solution $U$ of BSDE$^\circ$, $U^{\tau-}$ is a solution of the $(\mathbb{G},\mathbb{Q})$ BSDE$^X$. We get in particular the well-posedness of the BSDE$^X$ and of the corresponding counterparty risk model.

Condition (A) of \cite{CS1,CS2} presents a significant progress regarding to the counterparty risk literature based mainly on the more rigid immersion assumption. For the first time via condition (A), the fundamental role of the filtration behind the condition $\Delta_\tau V=0$ is revealed. On the practical side, condition (A) makes a complex BSDE with random horizon and endogenous terminal condition, to be solved by a classical BSDE with constant horizon and exogenous terminal condition (a useful property for numerical implementation).  The work in \cite{CS1,CS2} is worth an extension. Our eventual aim is to allow the methodology of \cite{CS1,CS2} applied in models where the local martingale pricing measures are replaced by local martingale deflators (pricing kernels). The first step towards this extension is a general formulation of condition (A). Actually condition (A) can be expressed in term of local martingale deflator. Let $\mathsf{p}$ be a $(\mathbb{F},\mathbb{Q})$ local martingale.
\begin{quote}\itshape
(\textbf{A\texttt{"}})\ For an $(\mathbb{F},\mathbb{Q})$ semimartingale $X$, $X^{\tau-}$ possesses a $(\mathbb{G},\mathbb{Q})$ local martingale deflator $\mathsf{q}$, whenever $\mathsf{p}$ is an $(\mathbb{F},\mathbb{Q})$ local martingale deflator for $X$. 
\end{quote}
In the case of condition (A), the above condition is valid for $\mathsf{p}=\frac{d\mathbb{P}}{d\mathbb{Q}}$ and $\mathsf{q}\equiv 1$. Condition(A\texttt{"}) leads straightforwardly to the deflator problem. As shown previously, once the correspondence $\mathsf{p}$-$\mathsf{q}$ is established (that we will do in this paper), the methodology of \cite{CS1,CS2} will be applicable to solve the pricing BSDEs.

This work is closely related to the no-arbitrage condition of the first kind (cf. \cite{K2010,K2012,KK, ST, Song-takaoka}). It integrates also with the literature of the insider trading problem (cf. \cite{AFK, ACDJ, AIF, FJS, GP,imkeller, Songdrift, zwierz}). Another observation is the striking resemblance between the formula of $\mathbb{F}$ reductions of $\mathbb{G}$ local martingales (cf. section \ref{Freduction}) and the formula of absolutely continuous probability changes (cf. \cite[Theorem 12.18]{HWY}). This observation calls up the approach by Girsanov's theorem of the enlargement of filtration problems presented in \cite{SongThesis, Song-local-solution, yoeurp}. The fundamental relation between $\mathbb{G}$-deflator for $S^{\tau}$ and the multiplicative decompositions of Az\'ema's supermartingale has been previously pointed out in \cite{FJS} in the continuous case, based on the result in \cite{NY}, and in \cite{AFK} in the discontinuous case, based on the paper \cite{K2014}. Nevertheless, it is to be noticed that the multiplicative formulas in \cite{K2014, NY} can not serve the deflator construction of $S^{{\tau}-}$, because of the discontinuous situation and of the predictable nature of the problem. Recall that this work applies the multiplicative formula of \cite{jacod}.

\

\subsection{Notations and conventions}

We refer to \cite{HWY, jacod} for semimartingale calculus. For a process $X$, we make the convention $X_{0-}=X_{0}$ so that the jump $\Delta_0X=0$. We use ${^{o\cdot\mathbb{F}}}\!\bullet, {^{p\cdot\mathbb{F}}}\!\bullet$ to denote the optional and the predictable projections with respect to a filtration $\mathbb{F}$, as well as $\bullet{^{o\cdot\mathbb{F}}}, \bullet{^{p\cdot\mathbb{F}}}$ for the corresponding dual projections. The stochastic integral will be denoted notably by \twoprimes{$_\centerdot$} (ex. $\int_0^t H_s dX_s = H{_\centerdot}X_t$ or $\int_0^t K_sH_s dX_s = KH{_\centerdot}X_t$). The stochastic integral $\int_0^t H_s dX_s$ is always supposed to be computed on $(0,t]$ so that $H{_\centerdot}X_0=0$. We consider deterministic as well as random intervals, all being denoted by the usual bracket system. A random interval such as $[S,T]=\{(s,\omega)\in \mathbb{R}_+\times\Omega: S(\omega)\leq s\leq T(\omega)\}$ is a subset in $\mathbb{R}_+\times\Omega$. For a subset $A\subset \Omega$, $\mathbb{R}_+\times A$ also is a random interval. For simplicity, we denote the intersection of the two random intervals by $A\cap[S,T]$. For any non negative random variable $T$, for any set $A$, $T_A$ denote the random variable $T\ind_A+\infty\ind_{A^c}$. The Dol\'ean-Dade exponential is denoted by \twoprimes{$\mathcal{E}$}. We will apply the semimartingale calculus on a predictable random interval. We refer to section 8 of \cite{HWY} for details. For a c\`adl\`ag process $X$ and a non negative random variable $R$, the stopping just prior to $R$ is defined as follows $$
X^{R-}=X\ind_{[0,R)}+X_{R-}\ind_{[R,\infty)}
=X^R-\Delta_RX\ind_{\{0<R\}}\ind_{[R,\infty)}
$$ 
(cf. \cite[Chapitre VI n$^\circ5$]{DM2}). 

\

\section{Filtrations and no arbitrage condition}\label{setting}

We consider a stochastic basis $(\Omega,\mathcal{B},\mathbb{G},\mathbb{Q})$ where $(\Omega,\mathcal{B})$ is a measurable space, $\mathbb{Q}$ is a probability measure on this measurable space, and $\mathbb{G}=(\mathcal{G}_t)_{t\in\mathbb{R}_+}$ is a filtration of sub-$\sigma$-algebras of $\mathcal{B}$ satisfying the usual condition. Let ${\tau}$ be a $\mathbb{G}$ stopping time. We assume the existence of a filtration $\mathbb{F}=(\mathcal{F}_t)_{t\in\mathbb{R}_+}$ satisfying the following condition.
\begin{assumption}\label{reduction}
\textbf{Reduction condition.} The filtration $\mathbb{F}$ is contained in $\mathbb{G}$. For any $\mathbb{G}$ optional (resp. predictable) process $H$, there exists an $\mathbb{F}$ optional (resp. predictable) process $K$ such that $H\ind_{[0,{\tau})}=K\ind_{[0,{\tau})}$ (resp. $H\ind_{(0,{\tau}]}=K\ind_{(0,{\tau}]}$).  We call the process $K$ an $\mathbb{F}$ optional (resp. predictable) reduction of the process $H$. 
\end{assumption}

Note that, if $\mathbb{G}$ is the classical progressive enlargement of $\mathbb{F}$ with ${\tau}$, the above condition will be satisfied. Moreover, the two properties in this condition are not independent.  For complementary discussions on this condition, see \cite[Chapitre XX n$^\circ$75]{DM3} and \cite{CS1,Jeulin80}. Besides the enlargement of filtration setting $\mathbb{F}\subset \mathbb{G}$, another basic notion in this paper is the notion of deflators.

\bd\label{defdef}
Let $S=(S^i)_{1\leq i\leq d}$ be a $d$-dimensional semimartingale in the filtration $\mathbb{F}$. A real valued strictly positive process $Y$ is called a (strictly positive local martingale) deflator in $\mathbb{F}$ for $S$, if $Y$ and $YS=(YS^i)_{1\leq i\leq d}$ are $\mathbb{F}$ local martingales.
\ed

We define the same notions in the filtration $\mathbb{G}$ in an obvious way. Note that the deflator notion is closely linked with the notion $\mathtt{NA}_1$ of the no-arbitrage condition of the first kind. See for example \cite{AFK, K2010, K2012, KK, ST, Song-takaoka}.

\bl
A semimartingale $S=(S^i)_{1\leq i\leq d}$ with strictly positive components satisfies the $\mathtt{NA}_1$ condition in a given filtration (i.e. $\mathtt{NA}_1$ condition on $[0,a]$ for any positive real number $a$), if and only if $S$ possesses a deflator. 
\el

\section{Az\'ema's supermartingale, its zeros, its decompositions}\label{ZP}

The deflator problem depends on Az\'ema's supermartingale in a delicate way. This section is devoted to an inventory of properties needed in the next sections. We recall that Az\'ema's supermartingale $Z$ in $\mathbb{F}$ associated with a random time ${\tau}$ is ${^{o\cdot\mathbb{F}}}\!(\ind_{[0,{\tau})})$ the $\mathbb{F}$ optional projection of $\ind_{[0,{\tau})}$. Associated with Az\'ema's supermartingale $Z$, there are the following processes.
\ebe 
\item[.]
The process $Z_-$ and ${^{p\cdot\mathbb{F}}}\!(\ind_{[0,\tau]})=\ind_{[0]}+\ind_{(0,\infty)}Z_-$ (cf. \cite{Jeulin80} p.63).

\item[.]
The process $\widetilde{Z}={^{o\cdot\mathbb{F}}}\!(\ind_{[0,{\tau}]})$.

\item[.]
The process $\mathsf{A}$ to be the $\mathbb{F}$ optional dual projection of $\ind_{\{0<{\tau}\}}\ind_{[{\tau},\infty)}$ and ${\replacem}=Z+\mathsf{A}$ which is an $\mathbb{F}$ $\mathtt{BMO}$ martingale (cf. \cite[Chapitre XX n$^\circ$74]{DM3}). By \cite{Jeulin80}, $$
\widetilde{Z}=Z+\Delta \mathsf{A}
=Z+\Delta ({\replacem}-Z)
=Z_- +\Delta {\replacem}
$$
on $(0,\infty)$. 

\item[.]
The process $\mathsf{a}$ to be the $\mathbb{F}$ predictable dual projection of $\ind_{\{0<{\tau}\}}\ind_{[{\tau},\infty)}$ and ${\replacemdot}=Z+\mathsf{a}$ which is an $\mathbb{F}$ $\mathtt{BMO}$ martingale (cf. \cite[Lemme(5.17)]{Jeulin80}). 

\item[.] 
The process ${\replacegamma}={^{p\cdot\mathbb{F}}}\!Z$ ($\mathbb{F}$ predictable projection of $Z$) $={\replacemdot}_--\mathsf{a}=Z_--\Delta\mathsf{a}$.
 
\dbe
%
%
%
%
%
%
%
%
%
%
With the process ${\replacegamma}$, we consider the $\mathbb{F}$ predictable set $\mathtt{C}(\frac{1}{{\replacegamma}})$ introduced in \cite[(6,23)]{jacod} by the relation:$$
\mbox{\itshape  for an $\mathbb{F}$ stopping time $T$, $\frac{1}{{\replacegamma}}\ind_{(0,T]}$ is locally bounded, if and only if $[0,T]\subset \mathtt{C}(\frac{1}{{\replacegamma}})$}.
$$
The $\mathbb{F}$ predictable set $\mathtt{C}(\frac{1}{{\replacegamma}})$ is introduced to have the predictable multiplicative decomposition of $Z$ defined in \cite[Theorem (6.31) and Exercice 6.10]{jacod}.

\bl\label{multidecomp}
There exists a non negative local martingale $L$ on $\mathtt{C}(\frac{1}{{\replacegamma}})$ and a predictable non negative non increasing process $D$ (on $\mathbb{R}_+$) such that $Z=LD$ on $\mathtt{C}(\frac{1}{{\replacegamma}})$. We have $L_0=1, D_0=Z_0$, and on $\mathtt{C}(\frac{1}{{\replacegamma}})$, $$
\dcb
\{L>0\}=\{0\}\cup\{Z>0\},\
\{D>0\}=\{Z_0>0\}.
\dce
$$
For an $\mathbb{F}$ stopping time $T$ such that $[0,T]\subset \mathtt{C}(\frac{1}{{\replacegamma}})$,$$
\dcb
L^T=\mathcal{E}[\frac{1}{{\replacegamma}}\ind_{(0,T]}{_\centerdot}{\replacemdot}],\
D^T=\frac{Z_0}{\mathcal{E}[\frac{1}{{\replacegamma}}\ind_{(0,T]}{_\centerdot}\mathsf{a}]}.
\dce
$$
\el

\

\subsection{The vanishing times and the sets of positive values}

Whether or not the processes $Z, Z_-, {\replacegamma}$ take positive values constitutes an important technical point in the following sections. This bring us to consider the sets $\{Z>0\}, \{Z_->0\}, \{{\replacegamma}>0\}$ and $\mathtt{C}(\frac{1}{{\replacegamma}})$. These sets are all random intervals starting from zero (included) up to a stopping time (included or not). To see the exact situation of them, we introduce the following stopping times (the vanishing times).
\ebe
\item[.] 
$\zeta=\inf\{s:Z_s=0 \mbox{ or } Z_{s-}=0\}$, and, for $n\in\mathbb{N}_+$, $\zeta_n=\inf\{s:Z_s\leq \frac{1}{n}\}$. 

\item[.]
$\eta=\zeta_{\{0<\zeta<\infty,Z_{\zeta-}>0,{\replacegamma}_{\zeta}>0\}}$.

\item[.]
$\dot{\eta}=\zeta_{\{\forall k, \zeta_k<\zeta\}}$. 

\item[.]
$\ddot{\eta}=\zeta_{\{0<\zeta<\infty,Z_{\zeta-}>0,{\replacegamma}_\zeta=0\}}$.

\item[.]
${\replacedotzeta}=\inf\{s:{\replacegamma}_s=0\}$.
\dbe

\bl\label{Z>0}
$\{Z> 0\}=\lc0,\zeta\lc$, $\zeta=\sup_n\zeta_n$ and ${\replacedotzeta}=\zeta$.
\el

\begin{proof}
The first properties are deduced from \cite[Chapitre VI $n^\circ$17]{DM2}. By the proof of \cite[Corollaire(6.28)]{jacod} ${\replacedotzeta}\geq \zeta$. We check also that ${\replacegamma}\ind_{\rc \zeta,\infty\lc}\leq Z_-\ind_{\rc\zeta,\infty\lc}=0$. Hence, ${\replacedotzeta}=\zeta$. 
\end{proof}

\bl\label{Z->0}
$\{Z_->0\}=\{Z_0>0\}\cap(\cup_n\lc0,\zeta_n\rc)$. 
\el

\begin{proof}
By Lemma \ref{Z>0}, $\{Z_0=0\}\cap\{Z_->0\}=\emptyset$ and $Z_-\ind_{\rc\zeta,\infty\lc}=0$. Hence, the interval $\{Z_->0\}$ writes as $\{Z_0>0\}\cap (\lc0,\zeta\lc\cup\lc\zeta_{\{0<\zeta<\infty, Z_{\zeta-}>0\}}\rc)$. As $\zeta=\sup_n\zeta_n$, $\lc0,\zeta\lc\subset\cup_n\lc0,\zeta_n\rc$. On the other hand, if $0<\zeta<\infty, Z_{\zeta-}>0$, by \cite[Lemma 3.1]{JS1} $\inf_{0\leq s<\zeta}Z_s>\frac{1}{n}$ for some $n>0$ so that $\zeta\leq \zeta_n$. We conclude $\{Z_->0\}\subset\cup_n\lc0,\zeta_n\rc$ on the set $\{Z_0>0\}$. The inverse inclusion is obvious. The lemma is proved.
\end{proof}

\bl\label{ttC}
$\{0<\zeta<\infty,Z_{\zeta-}=0\}=\{\dot{\eta}<\infty\}$, i.e., $\dot{\eta}=\zeta_{\{0<\zeta<\infty,Z_{\zeta-}=0\}}$. The random times $\dot{\eta},\ddot{\eta}$ are $\mathbb{F}$ predictable stopping times. The identities $\cup_n\lc 0,\zeta_n\rc=\lc0,\zeta\rc\setminus[\dot{\eta}]$ and $\{{\replacegamma}>0\}=\{Z_0>0\}\cap(\lc0,\zeta\rc\setminus (\lc\dot{\eta}\rc\cup\lc\ddot{\eta}\rc))$ hold.
\el

\begin{proof}
By definition, if $\dot{\eta}<\infty$, we have $0<\zeta=\dot{\eta}<\infty$ and, because of $\zeta_{n}<\zeta, \zeta=\sup_n\zeta_n$, $Z_{\zeta-}=\lim_nZ_{\zeta_n}=0$. On the other hand, if $0<\zeta_n<\infty$, we have $Z_{\zeta_n-}\geq\frac{1}{n}>0$ so that $\zeta_n<\zeta$ on the set $\{0<\zeta<\infty,Z_{\zeta-}=0\}$. This proves the first part of the lemma. The predictability of $\dot{\eta}$ is shown in \cite[proof of Theorem 9.41]{HWY}. By Lemma \ref{Z>0}, $Z_{-}= {\replacegamma}=0$ on $\rc\zeta,\infty\lc$. As ${\replacedotzeta}=\zeta$, $$
\lc\ddot{\eta}\rc=\{Z_->0,{\replacegamma}=0\},
$$
which is predictable. The identity $\cup_n\lc 0,\zeta_n\rc=\lc0,\zeta\rc\setminus[\dot{\eta}]$ is deduced directly from the definition of $\dot{\eta}$. By Lemma \ref{Z>0}, $\{Z_0=0\}\cap\{{\replacegamma}>0\}=\emptyset$ and ${\replacegamma}\ind_{\rc\zeta,\infty\lc}=0$. Hence, the interval $\{{\replacegamma}>0\}$ writes as $$
\{Z_0>0\}\cap (\lc0,\zeta\lc\cup\lc\zeta_{\{0<\zeta<\infty, {\replacegamma}_{\zeta}>0\}}\rc)
=
\{Z_0>0\}\cap (\lc0,\zeta\rc\setminus (\lc\dot{\eta}\rc\cup\lc\ddot{\eta}\rc))
$$ 
\end{proof}

\brem
The stopping time $\eta$ will be studied in Lemma \ref{nmart}. It will be proved that  $\eta\in\mathtt{C}(\frac{1}{{\replacegamma}})$ on $\{0<\eta<\infty\}$. 
\erem

Combining Lemma \ref{Z->0} and Lemma \ref{ttC} with \cite[(6.24) and (6.28)]{jacod}, we obtain the next lemma.

\bl\label{Cgamma}
We have $$
\mathtt{C}(\frac{1}{{\replacegamma}})=(\cup_n\lc0,\zeta_n\rc)\setminus \lc\ddot{\eta}\rc
=\lc0,\zeta\rc\setminus (\lc\dot{\eta}\rc\cup\lc\ddot{\eta}\rc)
$$
and $\{Z_0>0\}\cap\mathtt{C}(\frac{1}{{\replacegamma}})=\{{\replacegamma}>0\}$. There exists a non decreasing sequence of $\mathbb{F}$ stopping times $(S_n)_{n\in\mathbb{N}^*}$ such that $\mathtt{C}(\frac{1}{{\replacegamma}})=\cup_n\lc0,S_n\rc$ and, for any $n\in\mathbb{N}$, $\frac{1}{{\replacegamma}}\leq n$ on $\rc0,S_n\rc$.
\el

\subsection{The jumps $\Delta{\replacemdot},\Delta\mathsf{a}$ outside of $\mathtt{C}(\frac{1}{{\replacegamma}})$ and the positivity of $Z_{{\tau}-}$}

\bl\label{asupport}
$0=\Delta_{\dot{\eta}}{\replacemdot}=\Delta_{\dot{\eta}}\mathsf{a}$ on $\{0<\dot{\eta}<\infty\}$. $\Delta_{\ddot{\eta}}\mathsf{a}=Z_{\ddot{\eta}-}>0$ on $\{0<\ddot{\eta}<\infty\}$. $\cup_n\lc0,\zeta_n\rc=\mathtt{C}(\frac{1}{{\replacegamma}})\cup\lc\ddot{\eta}\rc$ is a support of the random measure $d\mathsf{a}_s$, that implies ${\tau}\in\cup_n\lc0,\zeta_n\rc$ almost surely. As a consequence, $Z_{{\tau}-}>0, D_{\tau-}>0$ on $\{{\tau}<\infty,Z_0>0\}$, and $L_{{\tau}-}>0$ on $\{{\tau}<\infty\}$ (the existence of $L_{{\tau}-}$ being proved in Lemma \ref{Llimit}).
\el

\begin{proof}
We have $0=\Delta_{\dot{\eta}}Z=\Delta_{\dot{\eta}}{\replacemdot}-\Delta_{\dot{\eta}}\mathsf{a}$ on $\{0<\dot{\eta}<\infty\}$. This means $\Delta_{\dot{\eta}}{\replacemdot}\ind_{\{0<\dot{\eta}<\infty\}}\in\mathcal{F}_{\dot{\eta}-}$. The stopping time $\dot{\eta}$ being predictable, necessarily $\Delta_{\dot{\eta}}{\replacemdot}=0$ on $\{0<\dot{\eta}<\infty\}$ (cf. \cite[Theorem 4.41]{HWY}). The second assertion of the lemma is the consequence of $\Delta_{\ddot{\eta}}\mathsf{a}=Z_{\ddot{\eta}-}-{\replacegamma}_{\ddot{\eta}-}=Z_{\ddot{\eta}-}>0$ on $\{0<\ddot{\eta}<\infty\}$. The third assertion is because $\lc0,\zeta\rc$ is a support of $d\mathsf{a}_s$ and $\Delta_{\dot{\eta}}\mathsf{a}=0$. The last assertions are direct consequences of ${\tau}\in\cup_n\lc0,\zeta_n\rc$, of Lemma \ref{Z->0} and of $Z=LD$ on $\lc0,\zeta\lc$ (cf. also \cite[Lemme 0]{yorlemma}).
\end{proof}

\bl\label{Da=1}
For any $\mathbb{F}$ predictable stopping time $\sigma$, let $\sigma'=\sigma_{\{\Delta_\sigma\mathsf{a}=1,0<\sigma<\infty\}}$. Then, $\lc\sigma'\rc\subset\lc{\tau}\rc$ and $Z_{\sigma'}=0$, $Z_{\sigma'-}=1$ on $\{\sigma'<\infty\}$.
\el

\begin{proof}
Note that $\sigma'$ is $\mathbb{F}$ predictable. Applying \cite[Theorem 5.27]{HWY}, we obtain$$
\mathbb{E}[\ind_{\{0<{\tau}=\sigma<\infty\}}\ind_{\{\Delta_\sigma\mathsf{a}=1\}}\ind_{\{0<\sigma<\infty\}}]
=
\mathbb{E}[\Delta_\sigma\mathsf{a}\ind_{\{\Delta_\sigma\mathsf{a}=1\}}\ind_{\{0<\sigma<\infty\}}]
=
\mathbb{E}[\ind_{\{\Delta_\sigma\mathsf{a}=1\}}\ind_{\{0<\sigma<\infty\}}],
$$
which implies that $\{\Delta_\sigma\mathsf{a}=1,0<\sigma<\infty\}\subset\{0<{\tau}=\sigma<\infty\}$. Consequently, $$
Z_{\sigma'}\ind_{\{\sigma'<\infty\}}=\mathbb{P}[\sigma'<{\tau}|\mathcal{F}_{\sigma'}]\ind_{\{\sigma'<\infty\}}=0,\
Z_{\sigma'-}\ind_{\{\sigma'<\infty\}}=\mathbb{P}[\sigma'\leq {\tau}|\mathcal{F}_{\sigma'-}]\ind_{\{\sigma'<\infty\}}=\ind_{\{\sigma'<\infty\}}.
$$
\end{proof}

\subsection{The predictable multiplicative decomposition of $Z$ reconsidered}\label{multiplicativedecomposition}

In this section we review the predictable multiplicative decompositions of $Z$. We begin with some new and useful expressions of $L$ and of $D$ on $\mathtt{C}(\frac{1}{{\replacegamma}})$.  

\bl\label{newMdecomp}
On the set $\mathtt{C}(\frac{1}{{\replacegamma}})$, 
\begin{equation}
\dcb
\mathcal{E}(-\frac{1}{Z_-}{_\centerdot}\mathsf{a})\mathcal{E}(\frac{1}{{\replacegamma}}{_\centerdot}\mathsf{a})\equiv 1,\\
L=1+\ind_{\{Z_0>0\}}\frac{1}{Z_0}\mathcal{E}(-\frac{1}{Z_-}{_\centerdot}\mathsf{a})^{-1}{_\centerdot}{\replacemdot},\\
D=Z_0\mathcal{E}(-\frac{1}{Z_-}{_\centerdot}\mathsf{a}). 
\dce
\end{equation} 
\el

\begin{proof}
Notice that, if $Z_{0}=0$, nothing is to be proved.

Let $(S_n)_{n\in\mathbb{N}}$ be the sequence of $\mathbb{F}$ stopping times introduced in Lemma \ref{Cgamma}. Since $\forall n>0, (0,S_n]\subset\{Z_->0\}$, $\mathcal{E}(-\frac{1}{Z_-}{_\centerdot}\mathsf{a})$ is well-defined on the set $\mathtt{C}(\frac{1}{{\replacegamma}})$. Using the integration by parts formula on each of $(0,S_{n}]$, we have$$
\dcb
&&d\left(\mathcal{E}(-\frac{1}{Z_-}{_\centerdot}\mathsf{a})\mathcal{E}(\frac{1}{{\replacegamma}}{_\centerdot}\mathsf{a})\right)\\
&=&
\mathcal{E}(-\frac{1}{Z_-}{_\centerdot}\mathsf{a})_-\mathcal{E}(\frac{1}{{\replacegamma}}{_\centerdot}\mathsf{a})_-(-\frac{1}{Z_-})d\mathsf{a}
+
\mathcal{E}(-\frac{1}{Z_-}{_\centerdot}\mathsf{a})_-\mathcal{E}(\frac{1}{{\replacegamma}}{_\centerdot}\mathsf{a})_-\frac{1}{{\replacegamma}}d\mathsf{a}\\
&&
+
\mathcal{E}(-\frac{1}{Z_-}{_\centerdot}\mathsf{a})_-\mathcal{E}(\frac{1}{{\replacegamma}}{_\centerdot}\mathsf{a})_-(-\frac{1}{Z_-})\frac{1}{{\replacegamma}}d[\mathsf{a},\mathsf{a}]\\

&=&
\mathcal{E}(-\frac{1}{Z_-}{_\centerdot}\mathsf{a})_-\mathcal{E}(\frac{1}{{\replacegamma}}{_\centerdot}\mathsf{a})_-
\left((-\frac{1}{Z_-})+\frac{1}{{\replacegamma}}+(-\frac{1}{Z_-})\frac{1}{{\replacegamma}}\Delta\mathsf{a}\right)d\mathsf{a}\\
&=&
\mathcal{E}(-\frac{1}{Z_-}{_\centerdot}\mathsf{a})_-\mathcal{E}(\frac{1}{{\replacegamma}}{_\centerdot}\mathsf{a})_-
\frac{-{\replacegamma}+Z_--\Delta\mathsf{a}}{{\replacegamma} Z_-}d\mathsf{a}\\
&=&0.
\dce
$$
It yields that $\mathcal{E}(-\frac{1}{Z_-}{_\centerdot}\mathsf{a})\mathcal{E}(\frac{1}{{\replacegamma}}{_\centerdot}\mathsf{a})\equiv 1$ on the set $\{Z_{0}>0\}\cap \mathtt{C}(\frac{1}{{\replacegamma}})$. We conclude with Lemma \ref{multidecomp} that $$
D=\frac{Z_0}{\mathcal{E}(\frac{1}{{\replacegamma}}{_\centerdot}\mathsf{a})}
=
Z_0\mathcal{E}(-\frac{1}{Z_-}{_\centerdot}\mathsf{a})
$$
on the set $\mathtt{C}(\frac{1}{{\replacegamma}})$. Again on the set $\rc0,S_n\rc$, $$
\dcb
d{\replacemdot}=d(Z+\mathsf{a})=d(LD+\mathsf{a})
=
DdL+L_-dD+d\mathsf{a}
=
DdL,
\dce
$$
because ${\replacemdot}$ is a martingale. This implies that $\frac{1}{D}$ is $\mathsf{m}$ integrable on the set $\{Z_0>0\}\cap\lc0,S_n\rc$, and consequently,$$
L=1+\ind_{\{Z_0>0\}}\frac{1}{Z_0}\mathcal{E}(-\frac{1}{Z_-}{_\centerdot}\mathsf{a})^{-1}{_\centerdot}{\replacemdot}\ \mbox{ on $\lc0,S_n\rc$}.
$$
\end{proof}

The predictable multiplicative decomposition of the Az\'ema supermartingale $Z$ in Lemma \ref{multidecomp} is defined on the set $\mathtt{C}(\frac{1}{{\replacegamma}})$. In the following lemma we will extend its definition to the whole $\mathbb{R}_{+}$. Recall $(S_n)_{n\in\mathbb{N}^*}$ the non decreasing sequence of $\mathbb{F}$ stopping times introduced in Lemma \ref{Cgamma} with $\mathtt{C}(\frac{1}{{\replacegamma}})=\cup_n\lc0,S_n\rc$. 

\bl\label{Llimit}
The left limit process $L_{-}$ is well defined and is finite on $(0,\zeta]$ (especially at $\zeta$). We can extend the definitions of the processes $L, D$ from the set $\mathtt{C}(\frac{1}{{\replacegamma}})$ to the whole $\mathbb{R}_{+}$ by setting 
\begin{equation}\label{Lextension}
\dcb
L
&=& \lim_{n\rightarrow\infty}L^{S_{n}}
=L\ind_{\lc 0,\zeta\lc}+L_{\zeta-}\ind_{\{\forall n>0, S_{n}<\zeta<\infty\}}\ind_{\lc\zeta,\infty\lc},
\dce
\end{equation}
\begin{equation}\label{Dextension}
\dcb
D
&=& 
\lim_{n\rightarrow\infty}D^{S_{n}}\ind_{[0,\ddot{\eta})}\\
&=&
(D\ind_{\lc 0,\zeta\lc}+D_{\zeta-}\ind_{\{\forall n>0, S_{n}<\zeta<\infty\}}\ind_{\lc\zeta,\infty\lc}+D_{\zeta}\ind_{\{\exists n>0, S_{n}=\zeta<\infty\}}\ind_{\lc\zeta,\infty\lc})\ind_{[0,\ddot{\eta})}.
\dce
\end{equation}
The redefined process $L$ is an $\mathbb{F}$ local martingale, and the redefined process $D$ is an $\mathbb{F}$ predictable process with non increasing path and they satisfy the multiplicative decomposition identity $Z=LD$ on the whole $\mathbb{R}_+$. 
\el

\begin{proof}
We use the same notions $L, D$ to denote the processes on the set $\mathtt{C}(\frac{1}{{\replacegamma}})$ already defined in Lemma \ref{multidecomp}, as well as the redefined processes in the present lemma. There will be no risk of confusion, because the latters are extensions of the formers.

We show firstly that the expression (\ref{Lextension}) defining the extended process $L$ is meaningful. To this end, introduce $
\hat{L}=\ind_{\{Z_0=0\}}\ind_{[0]}+L\ind_{\lc0,\zeta\lc}
$
(a process defined on the whole $\mathbb{R}_+$). 
Notice that $$
\dcb
L\ind_{\lc0,S_n\rc}
&=&
\ind_{\{S_n=0\}}\ind_{[0]}+\ind_{\{S_n>0\}}L\ind_{\lc0,S_n\rc}\ind_{\lc0,\zeta\lc},\ \mbox{ because on $\{S_n>0\}\subset\{\zeta>0\}$, $\{L>0\}=\lc0,\zeta\lc$,}\\
&=&
\ind_{\{S_n=0\}}\ind_{[0]}\hat{L}+\ind_{\{S_n>0\}}\hat{L}\ind_{\lc0,S_n\rc}
=\hat{L}\ind_{\lc0,S_n\rc}.
\dce
$$
This proves that $\hat{L}$ is (also) an extension of the process $L$ defined in Lemma \ref{multidecomp}, because they coincide on $\mathtt{C}(\frac{1}{{\replacegamma}})$. For any pair of $\mathbb{F}$ stopping times $T\leq T'$, for any positive integer $n>0$, $$
\dcb
&&
\mathbb{E}[\hat{L}_{T'}\ind_{\{T'\in \lc0,S_n\rc\}}|\mathcal{F}_T]
=
\mathbb{E}[L_{T'}\ind_{\{T'\in \lc0,S_n\rc\}}|\mathcal{F}_T]\\
&\leq&
\mathbb{E}[L_{T'\wedge S_n}|\mathcal{F}_T]\ind_{\{T\in \lc0,S_n\rc\}}
\leq
L_{T\wedge S_n}\ind_{\{T\in \lc0,S_n\rc\}}
=
\hat{L}_{T}\ind_{\{T\in \lc0,S_n\rc\}},
\dce
$$
because $L^{S_n}$ is a non negative $\mathbb{F}$ local martingale, hence a supermartingale. Fatou's lemma implies that $\hat{L}\ind_{\mathtt{C}(\frac{1}{{\replacegamma}})}=\hat{L}$ is an $\mathbb{F}$ supermartingale. As $L$ coincides with $\hat{L}$ on $[0,\zeta)$, the supermartingale property of $\hat{L}$ implies that $L_{\zeta-}=\lim_{n\rightarrow \infty}\hat{L}_{S_{n}}$ exists on $\{\forall n>0, S_{n}<\zeta<\infty\}$ and is finite. The extended process $L$ in (\ref{Lextension}) is well-defined.

Note that the redefined process $L$ can also be written as
$$
L
=
L\ind_{\lc 0,\zeta\lc}+L_{\zeta-}\ind_{\{\forall n>0, S_{n}<\zeta<\infty\}}\ind_{\lc\zeta,\infty\lc}+L_{\zeta}\ind_{\{\exists n>0, S_{n}=\zeta, 0<\zeta<\infty\}}\ind_{\lc\zeta,\infty\lc},
$$
because on $\{\exists n>0, S_{n}=\zeta, 0<\zeta<\infty\}$, $L_{\zeta}=0$. We see then that the extended $L$ is precisely the extension introduced in \cite[(5.7)]{jacod} of the process $L$ defined in Lemma \ref{multidecomp}. By \cite[Lemme 5.17]{jacod}, the redefined $L$ is an $\mathbb{F}$ supermartingale. Actually, this redefined process $L$ is an $\mathbb{F}$ local martingale. Let $L=M-V$ be the Doob-Meyer decomposition of the redefined $L$ with $\mathbb{F}$ local martingale $M$ and non decreasing $\mathbb{F}$ predictable process $V$ with $V_{0}=0$. As $L^{S_{n}}$ is an $\mathbb{F}$ local martingale, $\ind_{[0,S_{n}]}{_\centerdot} V=0$. As $L$ is stopped at $\zeta$, $\ind_{(\zeta,\infty)}{_\centerdot} V=0$. It results from Lemma \ref{Cgamma} that $V=(\ind_{[0,\zeta]}-\ind_{\mathtt{C}(\frac{1}{{\replacegamma}})}){_\centerdot}V=\Delta_{\dot{\eta}\wedge \ddot{\eta}}V$. Note also that $\{\dot{\eta}\wedge \ddot{\eta}<\infty\}=\{\forall n>0, S_{n}<\zeta<\infty\}$ and $\zeta=\dot{\eta}\wedge \ddot{\eta}$ if $\dot{\eta}\wedge \ddot{\eta}<\infty$. Therefore,$$
(\Delta_{\dot{\eta}\wedge \ddot{\eta}}M-\Delta_{\dot{\eta}\wedge \ddot{\eta}}V)\ind_{\{\dot{\eta}\wedge \ddot{\eta}<\infty\}}
=
\Delta_{\dot{\eta}\wedge \ddot{\eta}}L\ind_{\{\dot{\eta}\wedge \ddot{\eta}<\infty\}}
=
\ind_{\{\forall n>0, S_{n}<\zeta<\infty\}}\Delta_{\zeta}L=0.
$$
As $V$ is an $\mathbb{F}$ predictable process and $\dot{\eta}\wedge \ddot{\eta}$ is an $\mathbb{F}$ predictable stopping time (cf. Lemma \ref{ttC}), the computation of the generalized conditional expectation (cf. \cite[Theorem 1.17]{HWY}) of $\Delta_{\dot{\eta}\wedge \ddot{\eta}}V$ gives
$$
\Delta_{\dot{\eta}\wedge \ddot{\eta}}V\ind_{\{\dot{\eta}\wedge \ddot{\eta}<\infty\}}
=
\mathbb{E}[\Delta_{\dot{\eta}\wedge \ddot{\eta}}V\ |\mathcal{F}_{\dot{\eta}\wedge \ddot{\eta}-}]\ind_{\{\dot{\eta}\wedge \ddot{\eta}<\infty\}}
=
\mathbb{E}[\Delta_{\dot{\eta}\wedge \ddot{\eta}}M\ |\mathcal{F}_{\dot{\eta}\wedge \ddot{\eta}-}]\ind_{\{\dot{\eta}\wedge \ddot{\eta}<\infty\}}=0
$$
(cf. \cite[Theorem 7.13]{HWY}). This proves $V\equiv 0$ so that the redefined $L$ is an $\mathbb{F}$ local martingale.

Consider the redefined process $D$, which is clearly non increasing. Since $D^{S_{n}}$ and $\ddot{\eta}$ are predictable, $D$ is an $\mathbb{F}$ predictable process. 

To verify the identity $Z=LD$, we check it immediately on $[0]\cup\lc0,\zeta\lc$. As for the identity on the interval $(0,\infty)\cap\lc\zeta,\infty\lc$, on the one hand,$$
\dcb
&&LD\ind_{\{0<\zeta<\infty\}}\ind_{\lc\zeta,\infty\lc}
=
LD\ind_{[0,\ddot{\eta})}\ind_{\{0<\zeta<\infty\}}\ind_{\lc\zeta,\infty\lc}\\
&=&
L_{\zeta}D_{\zeta}\ind_{\{\exists n>0, S_{n}=\zeta, 0<\zeta<\infty\}}\ind_{[0,\ddot{\eta})}\ind_{\lc\zeta,\infty\lc}
+
L_{\zeta-}D_{\zeta-}\ind_{\{\forall n>0, S_{n}<\zeta<\infty\}}\ind_{[0,\ddot{\eta})}\ind_{\lc\zeta,\infty\lc}\\
&=&
L_{\zeta-}D_{\zeta-}\ind_{\{\forall n>0, S_{n}<\zeta<\infty\}}\ind_{[0,\ddot{\eta})}\ind_{\lc\zeta,\infty\lc},\ \mbox{ because  $\forall n>0$, on $\{S_{n}=\zeta>0\}$, $L_{\zeta}=0$}, \\
&=&
L_{\dot{\eta}-}D_{\dot{\eta}-}\ind_{\{\dot{\eta}<\infty\}}\ind_{\lc\zeta,\infty\lc}
+
L_{\ddot{\eta}-}D_{\ddot{\eta}-}\ind_{\{\ddot{\eta}<\infty\}}\ind_{[0,\ddot{\eta})}\ind_{\lc\zeta,\infty\lc}\\
&=&
Z_{\dot{\eta}-}\ind_{\{\dot{\eta}<\infty\}}\ind_{\lc\zeta,\infty\lc}
=
0=Z\ind_{\{0<\zeta<\infty\}}\ind_{\lc\zeta,\infty\lc},\\
\dce
$$
(cf. the first part of the proof of Lemma \ref{ttC} for $L_{\dot{\eta}-}D_{\dot{\eta}-}\ind_{\{\dot{\eta}<\infty\}}=Z_{\dot{\eta}-}\ind_{\{\dot{\eta}<\infty\}}=0$) and, on the other hand, $\zeta=0$ implies $D_{0}=Z_{0}=0$ and $\forall n>0, S_{n}=0$ so that $$
LD\ind_{\{0=\zeta\}}\ind_{\rc\zeta,\infty\lc}
=
L^0D^0\ind_{\{0=\zeta\}}\ind_{\rc\zeta,\infty\lc}
=0=Z\ind_{\{0=\zeta\}}\ind_{\rc\zeta,\infty\lc}.
$$
The lemma is proved.
\end{proof}

As usual, we write$$
\mathcal{E}(-\ind_{\{Z_{-}>0\}}\frac{1}{Z_-}{_\centerdot}\mathsf{a})_{t}
:=
\exp(-\ind_{\{Z_{-}>0\}}\frac{1}{Z_-}{_\centerdot}\mathsf{a}_{t})\prod_{0<s\leq t, Z_{s-}>0}(1-\frac{1}{Z_{s-}}\Delta_{s}\mathsf{a})e^{\frac{1}{Z_{s-}}\Delta_{s}\mathsf{a}},
\ t\geq 0.
$$

\bl\label{rearrange}
The redefined process $D$ in (\ref{Dextension}) satisfies
\begin{equation}\label{lastdef}
\dcb
D 
=
Z_0\mathcal{E}(-\ind_{\{Z_{-}>0\}}\frac{1}{Z_-}{_\centerdot}\mathsf{a})^{\dot{\eta}-}
=
Z_0\mathcal{E}(-\ind_{\{Z_{-}>0\}}\frac{1}{Z_-}{_\centerdot}\mathsf{a}).
\dce
\end{equation}
\el

\begin{proof}
According to lemma \ref{newMdecomp}, $$
\lim_{n\rightarrow\infty}D^{S_{n}}
=
\lim_{n\rightarrow\infty}Z_0\mathcal{E}(-\frac{1}{Z_-}{_\centerdot}\mathsf{a})^{S_{n}}
=
Z_0\mathcal{E}(-\ind_{\{Z_{-}>0\}}\frac{1}{Z_-}{_\centerdot}\mathsf{a})^{\dot{\eta}\wedge \ddot{\eta}-}\
(=
Z_0\mathcal{E}(-\ind_{\mathtt{C}(\frac{1}{{\replacegamma}})}\frac{1}{Z_-}{_\centerdot}\mathsf{a})).
$$
This yields that the redefined process $D$ satisfies$$
D 
=
Z_0\mathcal{E}(-\ind_{\{Z_{-}>0\}}\frac{1}{Z_-}{_\centerdot}\mathsf{a})^{\dot{\eta}\wedge \ddot{\eta}-}\ind_{[0,\ddot{\eta})} 
=
Z_0\mathcal{E}(-\ind_{\{Z_{-}>0\}}\frac{1}{Z_-}{_\centerdot}\mathsf{a})^{\dot{\eta}-}\ind_{[0,\ddot{\eta})}
$$
(which can be checked separately on $\{\ddot{\eta}=\infty\}$ and on $\{\ddot{\eta}<\infty\}\subset \{\dot{\eta}=\infty\}$).
Notice that, on the set $\{0<\ddot{\eta}<\infty\}$, $\Delta_{\ddot{\eta}}(-\ind_{\{Z_{-}>0\}}\frac{1}{Z_-}{_\centerdot}\mathsf{a})=-1$ (cf. Lemma \ref{asupport}) so that $\mathcal{E}(-\ind_{\{Z_{-}>0\}}\frac{1}{Z_-}{_\centerdot}\mathsf{a})\ind_{[0,\ddot{\eta})}=\mathcal{E}(-\ind_{\{Z_{-}>0\}}\frac{1}{Z_-}{_\centerdot}\mathsf{a})$ (cf. \cite[Theorem 9.41]{HWY}). Therefore, the above identity becomes
$$
D 
=
Z_0\mathcal{E}(-\ind_{\{Z_{-}>0\}}\frac{1}{Z_-}{_\centerdot}\mathsf{a})^{\dot{\eta}-}
=
Z_0\mathcal{E}(-\ind_{[0,\dot{\eta})}\ind_{\{Z_{-}>0\}}\frac{1}{Z_-}{_\centerdot}\mathsf{a})
=
Z_0\mathcal{E}(-\ind_{\{Z_{-}>0\}}\frac{1}{Z_-}{_\centerdot}\mathsf{a})
$$
(cf. Lemma \ref{Z->0} and Lemma \ref{ttC}).
\end{proof}

We can deduce from the multiplicative decomposition of Lemma \ref{Llimit} other multiplicative decompositions. We begin with a multiplicative decomposition for the random measure $d\mathsf{a}$.

\bl\label{daLD}
With the redefined processes $L,D$ in Lemma \ref{Llimit}, we have $d\mathsf{a}=-\ind_{\mathtt{C}(\frac{1}{{\replacegamma}})\cup[\ddot{\eta}]}L_-dD=-\ind_{[0,\zeta]}L_-dD$.
\el

\begin{proof}
By the integration by parts formula applied on the identity $Z=LD$ on the intervals $\lc0,S_n\rc$, where $(S_n)_{n\in\mathbb{N}^*}$ denotes the non decreasing sequence of $\mathbb{F}$ stopping times introduced in Lemma \ref{Cgamma}, we prove the lemma on the set $\mathtt{C}(\frac{1}{{\replacegamma}})=\cup_n\lc0,S_n\rc$. For the rest, it is the consequence of Lemma \ref{Cgamma}, of Lemma \ref{asupport} and of the following relations (for the redefined process $D$):$$
0<\Delta_{\ddot{\eta}}\mathsf{a}
=
Z_{\ddot{\eta}-}
=
L_{\ddot{\eta}-}D_{\ddot{\eta}-}
=
-L_{\ddot{\eta}-}\Delta_{\ddot{\eta}}D
$$
on $\{0<\ddot{\eta}<\infty\}$, and $\Delta_{\dot{\eta}}\mathsf{a}=0=\Delta_{\dot{\eta}}D$ on $\{0<\dot{\eta}<\infty\}$. 
\end{proof}

As a corollary of the identity $Z=LD$, we obtain also multiplicative decompositions of $Z_-$ and of ${\replacegamma}$ on the whole $\mathbb{R}_+$.

\bcor\label{Z-decompos}
With the redefined processes $L,D$ in Lemma \ref{Llimit}, $Z_-=L_-D_-$ and ${\replacegamma}=L_-D$ on $\mathbb{R}_+$.
\ecor

\subsection{Complements to the vanishing times and the sets of positive values}

By Lemma \ref{ttC}, the stopping times $\dot{\eta},\ddot{\eta}$ are $\mathbb{F}$ predictable stopping times. In contrast, the stopping time $\eta$ can never be a genuine $\mathbb{F}$ predictable stopping time, as shown in the next lemma (which is the counterpart in this paper of \cite[Lemma 3.5]{AFK}).  

\bl\label{nmart}
The stopping time $\eta$ coincides with $\zeta_{\{0<\zeta<\infty,Z_{\zeta-}>0,D_{\zeta}>0\}}$ (with the rearranged process $D$). On the set $\{0<\eta<\infty\}$, $\eta\in\mathtt{C}(\frac{1}{{\replacegamma}})$. Let $\replacemathsfd$ be the $\mathbb{F}$ predictable dual projection of $\ind_{\{0<\eta\}}\ind_{[\eta,\infty)}$ and $\mathsf{n}:=\mathcal{E}(-\replacemathsfd)^{-1}\ind_{[0,\eta)}$.
Then, The jump process $\Delta \replacemathsfd<1$ on the whole $\mathbb{R}_+$. The process $\mathsf{n}=\mathcal{E}(-\replacemathsfd)^{-1}\ind_{[0,\eta)}$ is well-defined and is an $\mathbb{F}$ local martingale.
\el

\begin{proof}
The first assertion is the consequence of the following equivalent relations :$$
\dcb
&&\{0<\zeta<\infty,Z_{\zeta-}>0,{\replacegamma}_{\zeta}>0\}\\
&=&
\{0<\zeta<\infty,L_{\zeta-}D_{\zeta-}>0,L_{\zeta-}D_{\zeta}>0,\zeta\in\mathtt{C}(\frac{1}{{\replacegamma}})\}\
\mbox{(cf. Corollary \ref{Z-decompos})}\\
&=&
\{0<\zeta<\infty,L_{\zeta-}D_{\zeta-}>0,D_{\zeta}>0,\zeta\in\mathtt{C}(\frac{1}{{\replacegamma}})\}\\
&=&
\{0<\zeta<\infty,Z_{\zeta-}>0,D_{\zeta}>0\}\ \mbox{because of Lemma \ref{Cgamma} and of the definition (\ref{lastdef}).}
\dce
$$
Note that, when $\eta<\infty$, $\dot{\eta}=\infty$ and $\ddot{\eta}=\infty$, proving the second assertion (cf. Lemma \ref{Cgamma}).  

Let $\sigma$ be an $\mathbb{F}$ predictable stopping time such that $\Delta_\sigma \replacemathsfd=1$ on $\{\sigma<\infty\}$. By \cite[Theorem 5.27]{HWY}, $
\mathbb{E}[\ind_{\{0<\eta=\sigma\}}|\mathcal{F}_{\sigma-}]=1
$
on $\{\sigma<\infty\}$, which implies $\{0<\sigma<\infty\}=\{0<\sigma=\eta<\infty\}$. We can make two computations on the set $\{0<\sigma<\infty\}$. On the one hand,$$
\dcb
L_{\sigma-}D_{\sigma-}
=
Z_{\sigma-}=\mathbb{E}[-\Delta_{\sigma}Z|\mathcal{F}_{\sigma-}]
=
\mathbb{E}[-\Delta_{\sigma}{\replacemdot}+\Delta_{\sigma}\mathsf{a}|\mathcal{F}_{\sigma-}]
=
\Delta_{\sigma}\mathsf{a}
=
-L_{\sigma-}\Delta_{\sigma}D,
\dce
$$
by Lemma \ref{daLD}, which yields $L_{\sigma-}D_{\sigma}=0$. On the other hand, $Z_{\sigma-}=Z_{\eta-}>0$ and $L_{\sigma-}D_{\sigma}={\replacegamma}_{\sigma}={\replacegamma}_{\eta}>0$. The two computations lead to contradictory results. We have proved that $\Delta_\sigma \replacemathsfd=1$ on $\mathbb{R}_{+}$. As a consequence, $\mathcal{E}(-\replacemathsfd)$ does not vanish (cf. \cite[Theorem 3.33 and Lemma 9.40]{HWY}).

To check the martingale property of $\mathsf{n}$, we apply the integration by parts formula on the two sides of the identity: $
1=\frac{\mathcal{E}(-\replacemathsfd)}{\mathcal{E}(-\replacemathsfd)}.
$
We obtain$$
d\replacemathsfd= \mathcal{E}(-\replacemathsfd)d(\frac{1}{\mathcal{E}(-\replacemathsfd)}).
$$
Consequently,$$
d\mathsf{n}=\ind_{\{\eta>0\}}\ind_{(0,\eta]}\frac{1}{\mathcal{E}(-\replacemathsfd)}d\replacemathsfd
-
\ind_{\{\eta>0\}}\frac{1}{\mathcal{E}(-\replacemathsfd)}d(\ind_{[\eta,\infty)})
=
-\frac{1}{\mathcal{E}(-\replacemathsfd)}d(\ind_{\{\eta>0\}}\ind_{[\eta,\infty)}-\replacemathsfd),
$$
which is an $\mathbb{F}$ local martingale.
\end{proof}

The vanishing sets $\{Z=0\}, \{Z_-=0\},\{{\replacegamma}=0\}$ are intrinsic features of the random time $\tau$, as shown in the next lemma.

\bl\label{chgpas}
Let $\mathbb{Q}'$ be a probability measure equivalent to $\mathbb{Q}$. Let $Z'$ and ${\replacegamma}'$ denote respectively the Az\'ema supermartingale of ${\tau}$ and its predictable projection computed under $\mathbb{Q}'$. Then, $\{Z'>0\}=\{Z>0\}$, $\{Z'_->0\}=\{Z_->0\}$ and $\{{\replacegamma}'>0\}=\{{\replacegamma}>0\}$. Consequently, if $\zeta',\eta', \dot{\eta}', \ddot{\eta}'$ denote the counterparts under $\mathbb{Q}'$ of $\zeta,\eta,\dot{\eta}, \ddot{\eta}$, we have $\zeta'=\zeta, \eta'=\eta, \dot{\eta}'=\dot{\eta}, \ddot{\eta}'=\ddot{\eta}$.
\el

\begin{proof}
For any $\mathbb{F}$ predictable stopping time $\sigma$, by the equivalence between $\mathbb{Q}$ and $\mathbb{Q}'$, we have$$
\dcb
&&\{Z'_{\sigma-}=0, 0<\sigma<\infty\}
=\{\mathbb{Q}'[\sigma\leq {\tau}|\mathcal{F}_{\sigma-}]=0, 0<\sigma<\infty\}\\
&=&
\{\mathbb{Q}[\sigma\leq {\tau}|\mathcal{F}_{\sigma-}]=0, 0<\sigma<\infty\}
=
\{Z_{\sigma-}=0, 0<\sigma<\infty\}
\dce
$$ 
and
$$
\{Z'_{0-}=0\}
=\{\mathbb{Q}'[0< {\tau}|\mathcal{F}_{0}]=0\}
=
\{\mathbb{Q}[0< {\tau}|\mathcal{F}_{0}]=0\}
=
\{Z_{0-}=0\}.
$$
The section theorem (cf. \cite[Corollary 4.11]{HWY}) proves $\{Z'_-=0\}=\{Z_-=0\}$. The identities $\{Z'=0\}=\{Z=0\}$ and $\{{\replacegamma}'=0\}=\{{\replacegamma}=0\}$ can be proved similarly. Consequently, $\zeta'=\zeta$. The identities $\eta'=\eta, \dot{\eta}'=\dot{\eta}, \ddot{\eta}'=\ddot{\eta}$ come then from the explicit definition of these stopping times (cf. Lemma \ref{ttC}).
\end{proof}

\

\section{Jump compensation martingale at a stopping time and the corresponding orthogonal decomposition}\label{JRO}

Let $R$ be an $\mathbb{F}$ stopping time and let ${^{R}}\!\mathsf{v}=(\ind_{\{0<R\}}\ind_{[R,\infty)})^{p\cdot\mathbb{F}}$ (the $\mathbb{F}$ predictable dual projection) and ${^{R}}\!\mathsf{u}=\ind_{\{0<R\}}\ind_{[R,\infty)}-{^{R}}\!\mathsf{v}$. In this section we study the $\mathbb{F}$ local martingales of the form $X^{R-}$.

\bl\label{ortho}
For any $\mathbb{F}$ local martingale $X$, $X^{R-}$ also is an $\mathbb{F}$ local martingale, if and only if $X$ is orthogonal to ${^{R}}\!\mathsf{u}$, i.e., ${^{R}}\!\mathsf{u} X$ is an $\mathbb{F}$ local martingale.
\el

\begin{proof}
It is the consequence of the following computation for any finite $\mathbb{F}$ stopping time $\sigma$ localizing the local martingales below:
$$
\dcb
&&
\mathbb{E}[X^{R-}_\sigma\ind_{\{0<R\}}]
=
\mathbb{E}[X_\sigma\ind_{\{\sigma<R\}}]+\mathbb{E}[X_{R-}\ind_{\{0<R\leq\sigma\}}]\\
&=&
\mathbb{E}[X_\sigma(\ind_{\{0<R\}}-\ind_{\{0<R\leq \sigma\}})]+\mathbb{E}[\int_0^\sigma X_{s-} d{^{R}}\!\mathsf{v}_s]\\
&=&
\mathbb{E}[X_\sigma\ind_{\{0<R\}}]-\mathbb{E}[X_\sigma\ind_{\{0<R\leq \sigma\}}]+\mathbb{E}[X_{\sigma} {^{R}}\!\mathsf{v}_\sigma]
=
\mathbb{E}[X_0\ind_{\{0<R\}}]-\mathbb{E}[X_\sigma{^{R}}\!\mathsf{u}_\sigma].
\dce
$$
\end{proof}

\bethe\label{Rdecomp}
Let $R^\natural=R_{\{0<R<\infty,\Delta_R{^{R}}\!\mathsf{v}=1\}}$. Then, $[R^\natural]=\{\Delta{^{R}}\!\mathsf{v}=1\}$ so that $\ind_{\{\Delta{^{R}}\!\mathsf{v}=1\}}{_\centerdot}{^{R}}\!\mathsf{u}=0$ and $R^\natural$ is an $\mathbb{F}$ predictable stopping time. For any $\mathbb{F}$ predictable stopping time $\sigma$ such that $[\sigma]\subset[R]$, we have $[\sigma]\subset[R^\natural]$. The jump process $\Delta{^{R}}\!\mathsf{v}\leq 1$. Let $R^\flat=R_{\{0<R<\infty,\Delta_R{^{R}}\!\mathsf{v}<1\}}$. Then, $$
(\ind_{\{0<R^\flat\}}\ind_{[R^\flat,\infty)})^{p\cdot\mathbb{F}}
=
\ind_{\{\Delta{^{R}}\!\mathsf{v}<1\}}{_\centerdot}{^{R}}\!\mathsf{v}.
$$ 
Let $\kappa=\ind_{\{\Delta{^{R}}\!\mathsf{v}<1\}}\Delta{^{R}}\!\mathsf{v}$. Almost surely, the path of the process $\frac{1}{1-\kappa}$ is bounded on every bounded interval. Let $X$ be an $\mathbb{F}$ local martingale. Let $K$ denote the $\mathbb{F}$ predictable process such that $$
\mathbb{E}[\Delta_R X|\mathcal{F}_{R-}]\ind_{\{0<R<\infty\}}
=
K_R\ind_{\{0<R<\infty\}},
$$
and let ${\replaceHX}X=\frac{K}{1-\kappa}$. Then, ${\replaceHX}X$ is ${^{R}}\!\mathsf{u}$ integrable in $\mathbb{F}$ and $X-{\replaceHX}X{_\centerdot}{^{R}}\!\mathsf{u}$ is orthogonal to ${^{R}}\!\mathsf{u}$ in $\mathbb{F}$. 
\ethe

\begin{proof}
For any $\mathbb{F}$ predictable stopping time $\sigma$, applying \cite[Theorem 5.27]{HWY}$$
\mathbb{E}[\ind_{\{0<R=\sigma\}}\ind_{\{\Delta_\sigma{^{R}}\!\mathsf{v}=1\}}\ind_{\{\sigma<\infty\}}]
=
\mathbb{E}[\Delta_\sigma{^{R}}\!\mathsf{v}\ind_{\{\Delta_\sigma{^{R}}\!\mathsf{v}=1\}}\ind_{\{0<\sigma<\infty\}}]
=
\mathbb{E}[\ind_{\{\Delta_\sigma{^{R}}\!\mathsf{v}=1\}}\ind_{\{0<\sigma<\infty\}}],
$$
which implies that $\{\Delta_\sigma{^{R}}\!\mathsf{v}=1,0<\sigma<\infty\}\subset\{0<R=\sigma<\infty\}$. We have$$
\ind_{\{\Delta_\sigma{^{R}}\!\mathsf{v}=1,0<\sigma<\infty\}}{^{p\cdot\mathbb{F}}}\!(\ind_{[0,R)})_\sigma
=
\mathbb{E}[\ind_{\{\Delta_\sigma{^{R}}\!\mathsf{v}=1,0<\sigma<\infty\}}\ind_{\{\sigma<R\}}|\mathcal{F}_{\sigma-}]
=0.
$$
The section theorem (cf. \cite[Theorem 4.8]{HWY}) implies that $\{\Delta{^{R}}\!\mathsf{v}=1\}\cap\{{^{p\cdot\mathbb{F}}}\!(\ind_{[0,R)})>0\}=\emptyset$ or equivalently $\{\Delta{^{R}}\!\mathsf{v}=1\}\subset\{{^{p\cdot\mathbb{F}}}\!(\ind_{[R,\infty)})=1\}$, which is a subset in $[R,\infty)$ according to \cite[Lemma (4.3)]{Jeulin80}.
But clearly, $\{\Delta{^{R}}\!\mathsf{v}=1\}\subset[0,R]$. We conclude $$
[R^\natural]=\{\Delta{^{R}}\!\mathsf{v}=1\}
$$
which is $\mathbb{F}$ predictable, proving the first assertion. For any $\mathbb{F}$ predictable stopping time $\sigma$ such that $\{0<\sigma<\infty\}\cap[\sigma]\subset[R]$, $$
\Delta_\sigma{^{R}}\!\mathsf{v}\ind_{\{0<\sigma<\infty\}}
=
\mathbb{E}[\ind_{\{0<R=\sigma\}}|\mathcal{F}_{\sigma-}]\ind_{\{\sigma<\infty\}}
=
\ind_{\{0<\sigma<\infty\}}.
$$
This means $\Delta_\sigma{^{R}}\!\mathsf{v}=1$ on $\{0<\sigma<\infty\}$, or in other words, $[\sigma]\subset[R^\natural]$ proving the second assertion. For any $\mathbb{F}$ predictable stopping time $\sigma$ such that $\Delta_\sigma{^{R}}\!\mathsf{v}\geq 1+\epsilon$ (for some $\epsilon>0$), $$
\mathbb{E}[\ind_{\{R=\sigma\}}\ind_{\{\Delta_\sigma{^{R}}\!\mathsf{v}\geq 1+\epsilon\}}\ind_{\{0<\sigma<\infty\}}]
=
\mathbb{E}[\Delta_\sigma{^{R}}\!\mathsf{v}\ind_{\{\Delta_\sigma{^{R}}\!\mathsf{v}\geq 1+\epsilon\}}\ind_{\{0<\sigma<\infty\}}]
>
\mathbb{E}[\ind_{\{\Delta_\sigma{^{R}}\!\mathsf{v}\geq 1+\epsilon\}}\ind_{\{0<\sigma<\infty\}}],
$$
if the last expectation is strictly positive. But this is impossible, because $\ind_{\{R=\sigma\}}\leq 1$. Hence, $\Delta_\sigma{^{R}}\!\mathsf{v}\leq 1$ almost surely, proving the third assertion. For any bounded $\mathbb{F}$ predictable process $H$,$$
\dcb
\mathbb{E}[H_{R^\flat}\ind_{\{R^\flat<\infty\}}]
=
\mathbb{E}[H_{R}\ind_{\{\Delta_R{^{R}}\!\mathsf{v}<1\}}\ind_{\{0<R<\infty\}}]
=
\mathbb{E}[\int_0^\infty H_{s}\ind_{\{\Delta_s{^{R}}\!\mathsf{v}<1\}}d{^{R}}\!\mathsf{v}_s]
\dce
$$
proving the fourth assertion. Let us write$$
\frac{1}{1-\kappa}
=
\frac{1}{1-\kappa}\ind_{\{\kappa\leq \frac{1}{2}\}}
+
\frac{1}{1-\kappa}\ind_{\{\kappa> \frac{1}{2}\}}.
$$
The first term in the right hand side is bounded by $2$. The second term is a finite discrete thin process. We prove thus the fifth assertion.

As for the rest assertion, we can suppose that $\Delta_RX$ is integrable. Hence, $K_R\ind_{[R,\infty)}$ is a process with integrable total variation, which implies the integrability of the total variation of $K{_\centerdot}{^{R}}\!\mathsf{v}$. As $\frac{1}{1-\kappa}$ has bounded path, $
\frac{K}{1-\kappa}{_\centerdot}{^{R}}\!\mathsf{v}
$
is a finite $\mathbb{F}$ predictable c\`adl\`ag process with finite variation. It is hence $\mathbb{F}$ locally bounded (cf. \cite[Theorem 5.19]{HWY}). The ${^{R}}\!\mathsf{u}$ integrability of ${\replaceHX}X=\frac{K}{1-\kappa}$ is proved. For any finite $\mathbb{F}$ stopping time $\sigma$ localizing the different local martingales in the computation, using Yoeurp's lemma (cf. \cite[Examples 9.4]{HWY}) and the predictability of $R^\natural$, we have on the one hand,
$$
\dcb
&&\mathbb{E}[X_\sigma {^{R}}\!\mathsf{u}_\sigma]
=
\mathbb{E}[[X,{^{R}}\!\mathsf{u}]_\sigma]
=
\mathbb{E}[[X,\ind_{\{0<R\}}\ind_{[R,\infty)}-{^{R}}\!\mathsf{v}]_\sigma]\\
&=&
\mathbb{E}[[X,\ind_{\{0<R\}}\ind_{[R,\infty)}]_\sigma]
=
\mathbb{E}[\Delta_R X\ind_{\{0<R\leq\sigma\}}]\\
&=&
\mathbb{E}[\Delta_R X\ind_{\{0<R^\natural\leq\sigma\}}]+\mathbb{E}[\Delta_R X\ind_{\{0<R^\flat\leq\sigma\}}]
=
\mathbb{E}[\Delta_R X\ind_{\{0<R^\flat\leq\sigma\}}].
\dce
$$
On the other hand, if $H$ denotes ${\replaceHX}X$,
$$
\dcb
&&\mathbb{E}[(H{_\centerdot}{^{R}}\!\mathsf{u}_\sigma) {^{R}}\!\mathsf{u}_\sigma]
=
\mathbb{E}[H{_\centerdot}[{^{R}}\!\mathsf{u},{^{R}}\!\mathsf{u}]_\sigma]
=
\mathbb{E}[H{_\centerdot}[{^{R}}\!\mathsf{u},\ind_{\{0<R\}}\ind_{[R,\infty)}]_\sigma]\\
&=&
\mathbb{E}[H_R(1-\Delta_R{^{R}}\!\mathsf{v})\ind_{\{0<R\leq\sigma\}}]
=
\mathbb{E}[\Delta_R X\frac{1}{1-\kappa_R}(1-\Delta_R{^{R}}\!\mathsf{v})\ind_{\{0<R\leq\sigma\}}]
=
\mathbb{E}[\Delta_R X\ind_{\{0<R^\flat\leq\sigma\}}].
\dce
$$
We obtain $\mathbb{E}[(X-H{_\centerdot}{^{R}}\!\mathsf{u})_\sigma {^{R}}\!\mathsf{u}_\sigma]=0$.
\end{proof}

\brem\label{Rdecomposition}
For any $\mathbb{F}$ local martingale $X$, let $$
{^{R}}\overline{X}=(X-{\replaceHX}X{_\centerdot}{^{R}}\!\mathsf{u})^{R-}
\mbox{ and } \overline{\Delta}_RX=\ind_{\{0<R\}}\Delta_R(X-{\replaceHX}X{_\centerdot}{^{R}}\!\mathsf{u}).
$$ 
Then, ${^{R}}\overline{X}, \ \overline{\Delta}_RX\ind_{\{0<R\}}\ind_{[R,\infty)}$ are $\mathbb{F}$ local martingales. The following decomposition for $X$ will be useful: $$
X^R 
=
{^{R}}\overline{X}+\overline{\Delta}_RX\ind_{\{0<R\}}\ind_{[R,\infty)}+ {\replaceHX}X{_\centerdot}{^{R}}\!\mathsf{u}. 
$$ 
\erem

\

\section{Deflators for a semimartingale with no jump at a stopping time}\label{DFJ}

In this section $R$ denotes an $\mathbb{F}$ stopping time. We use the notations ${^{R}}\!\mathsf{v},{^{R}}\!\mathsf{u}$ introduced in section \ref{JRO}. We now apply the results in section \ref{JRO} to give an answer to the question raised in \cite{AFK}
 
\bethe\label{ddotxi}
Let $S$ be an $\mathbb{F}$ semimartingale. Suppose that $S^{R-}$ has in $\mathbb{F}$ a deflator in exponential form $\mathcal{E}(\xi)$. 
Let $H={\replaceHX}\xi$. Then, $\mathcal{E}({^{R}}\overline{\xi}+H{_\centerdot}{^{R}}\!\mathsf{u})$
is also a deflator for $S^{R-}$. The $\mathbb{F}$ predictable process $\frac{1}{(1-H\Delta {^R}\!\mathsf{v})}$ is $\mathbb{F}$ locally bounded. The stochastic exponential $\mathcal{E}(\frac{1}{(1-H\Delta {^R}\!\mathsf{v})}{_\centerdot}{^{R}}\overline{\xi})$ is again an $\mathbb{F}$ deflator for $S^{R-}$.
\ethe

\begin{proof}
Recall the decomposition of $\xi$ by Remark \ref{Rdecomposition}:$$
\xi
=
{^{R}}\overline{\xi}+\overline{\Delta}_R\xi\ind_{\{0<R\}}\ind_{[R,\infty)}+H{_\centerdot}{^{R}}\!\mathsf{u}.
$$
Let us denote simply ${\replaceddotxi}$ for ${^{R}}\overline{\xi}$. As in \cite{Songdrift}, we consider the so-called structure condition (written componentwisely, obtained by the integration by parts formula):$$
S^{R-}+[\xi, S^{R-}]=\mbox{ $\mathbb{F}$ local martingale.}
$$
It is clear that the structure condition holds with ${\replaceddotxi}+H{_\centerdot}{^{R}}\!\mathsf{u}$, because $[\xi, S^{R-}]=[{\replaceddotxi}+H{_\centerdot}{^{R}}\!\mathsf{u}, S^{R-}]$. $\mathcal{E}({\replaceddotxi}+H{_\centerdot}{^{R}}\!\mathsf{u})$ will be a deflator for $S^{R-}$ whenever it is strictly positive. The positivity of $\mathcal{E}(\xi)$ implies $\Delta\xi>-1$. Hence,$$
\Delta({\replaceddotxi}+H{_\centerdot}{^{R}}\!\mathsf{u})
=
\Delta\xi >-1\  \mbox{ on $(0,R)$.}
$$
By the definition of $H$ (cf. Theorem \ref{Rdecomp}), $$
H_{R}\Delta_{R}{^{R}}\!\mathsf{u}\ind_{\{0<R<\infty\}}
=
\mathbb{E}[\Delta_{R}\xi|\mathcal{F}_{R-}]\ind_{\{\Delta_{R}{^{R}}\!\mathsf{u}\neq 0,0<R<\infty\}}>-1.
$$
We conclude that $\mathcal{E}({\replaceddotxi}+H{_\centerdot}{^{R}}\!\mathsf{u})>0$. 

We now address the second assertion of the theorem. Note that $H\Delta {^R}\!\mathsf{v}$ is the jump process of $H{_\centerdot}{^R}\!\mathsf{v}$. Let $\sigma$ be the d\'ebut $$
\sigma=\inf\{s> 0:H_s\Delta_s {^R}\!\mathsf{v}\geq 1\}
$$
which is an $\mathbb{F}$ predictable stopping time, because $\sigma\in\{H\Delta {^R}\!\mathsf{v}\geq 1\}$ if $\sigma<\infty$ (cf. \cite[Theorem 4.31]{HWY}). Because $\Delta\xi=\Delta {\replaceddotxi}-H\Delta {^R}\!\mathsf{v}>-1$ on $(0,R)$, if $0<\sigma<R$, we must have $
\Delta_{\sigma}{\replaceddotxi}>0.
$ 
As $\Delta{\replaceddotxi}=0$ on $[R,\infty)$, $\Delta_{\sigma}{\replaceddotxi}$ must be not negative on $\{0<\sigma<\infty\}$. On the other hand, a local martingale can not have a non negative non null jump at a finite predictable time. Necessarily $\Delta_{\sigma}{\replaceddotxi}=0$ on $\{0<\sigma<\infty\}$. This means that $\sigma\geq R$. But $\Delta {^R}\!\mathsf{v}=0$ on $(R,\infty)$ so that $\sigma=R$ on $\{0<\sigma<\infty\}$. By Theorem \ref{Rdecomp}, $\Delta_\sigma{^R}\!\mathsf{v}=1$ if $\sigma<\infty$. Let $K$ be the process in Theorem \ref{Rdecomp} associated with $\xi$. Let $J$ be a bounded $\mathbb{F}$ predictable process making everything in the computation integrable. 
$$
\dcb
&&\mathbb{E}[K_\sigma J_{\sigma}\ind_{\{0<\sigma<\infty\}}]\\
&=&\mathbb{E}[K_R J_{R}\ind_{\{0<R<\infty\}}\ind_{\{\sigma=R\}}]
=
\mathbb{E}[\Delta_R\xi J_{R}\ind_{\{0<R<\infty\}}\ind_{\{\sigma=R\}}]\\
&=&
\mathbb{E}[\Delta_\sigma\xi J_{\sigma}\ind_{\{0<\sigma<\infty\}}\ind_{\{\sigma=R^\natural\}}]
=
\mathbb{E}[\ \mathbb{E}[\Delta_\sigma\xi|\mathcal{F}_{\sigma-}]\  J_{\sigma}\ind_{\{0<\sigma<\infty\}}\ind_{\{\sigma=R^\natural\}}]=0,
\dce
$$
which implies that $K_\sigma=0$ on $\{0<\sigma<\infty\}$, and consequently $H_\sigma\Delta_\sigma {^R}\!\mathsf{v}=0$ if $\sigma<\infty$ (a contradiction to the definition of $\sigma$). We conclude $\sigma=\infty$. For $n\in\mathbb{N}$ let $$
\sigma_n=\inf\{s>0: H_s\Delta_s {^R}\!\mathsf{v}\geq 1-\frac{1}{n+2}\}.
$$
These $\sigma_n$ are $\mathbb{F}$ predictable stopping times and $\sigma_n\uparrow\infty$ (because, otherwise, $\sigma<\infty$). There exists then a non decreasing sequence $(\tilde{\sigma}_n)_{n\in\mathbb{N}}$ of $\mathbb{F}$ predictable stopping times such that $\tilde{\sigma}_n<\sigma_n$ and $\tilde{\sigma}_n\uparrow\infty$. We have thus shown that the $\mathbb{F}$ predictable process $\frac{1}{1-H\Delta {^R}\!\mathsf{v}}$ is $\mathbb{F}$ locally bounded.

Note that the structure condition is read as
$$
\dcb
\mbox{$\mathbb{F}$ local martingale}
&=&
S^{R-}+[\xi,S^{R-}]
=
S^{R-}+[{\replaceddotxi},S^{R-}]-[H{_\centerdot} {^R}\!\mathsf{v},S^{R-}]\\
&=&
(1-H\Delta {^R}\!\mathsf{v}){_\centerdot}S^{R-}+[{\replaceddotxi},S^{R-}]\
\mbox{(cf. \cite[Examples 9.4]{HWY})}.
\dce
$$
This is equivalent to
$$
\dcb
S^{R-}+[\frac{1}{(1-H\Delta {^R}\!\mathsf{v})}{_\centerdot}{\replaceddotxi},S^{R-}]
=\mbox{$\mathbb{F}$ local martingale.}
\dce
$$
Moreover,$$
\dcb
&&\Delta (\frac{1}{(1-H\Delta {^R}\!\mathsf{v})}{_\centerdot}{\replaceddotxi})
=
\frac{1}{(1-H\Delta {^R}\!\mathsf{v})}\Delta{\replaceddotxi}\ind_{[0,R)}\\
&=&
\frac{1}{(1-H\Delta {^R}\!\mathsf{v})}(\Delta\xi+H\Delta {^R}\!\mathsf{v})\ind_{[0,R)}
=
\frac{1}{(1-H\Delta {^R}\!\mathsf{v})}(\Delta\xi+1-(1-H\Delta {^R}\!\mathsf{v}))\ind_{[0,R)}
>-1.
\dce
$$
We have thus proved that $\mathcal{E}(\frac{1}{(1-H\Delta {^R}\!\mathsf{v})}{_\centerdot}{\replaceddotxi})$ is an $\mathbb{F}$ deflator for $S^{R-}$.
\end{proof}

\bcor\label{DY=0}
Let $S$ be an $\mathbb{F}$ semimartingale. Suppose that $S^{R-}$ has in $\mathbb{F}$ a deflator. Then, $S^{R-}$ has also an $\mathbb{F}$ deflator $Y$ such that $\Delta_RY=0$ on $\{0<R<\infty\}$.
\ecor


\

\section{Reduction from $\mathbb{G}$ into $\mathbb{F}$}\label{Freduction}

Recall the assumption that, for any $\mathbb{G}$ optional (resp. predictable) process $H$, there exists an $\mathbb{F}$ optional (resp. predictable) process $K$ such that $H\ind_{[0,{\tau})}=K\ind_{[0,{\tau})}$ (resp. $H\ind_{(0,{\tau}]}=K\ind_{(0,{\tau}]}$). We call the process $K$ an $\mathbb{F}$ optional (resp. predictable) reduction of the process $H$. 

\bl\label{rdps}
If a $\mathbb{G}$ optional process is strict positive on $[0,{\tau})$, its $\mathbb{F}$ optional reduction is strict positive on $[0,\zeta)$. If a $\mathbb{G}$ predictable process is strict positive on $(0,{\tau}]$, its $\mathbb{F}$ predictable reduction is strict positive on $\cup_n[0,\zeta_n]\setminus\{0\}$.
\el

\begin{proof}
Let $H$ be a $\mathbb{G}$ optional process, strict positive on $[0,{\tau})$. Let $K$ be an $\mathbb{F}$ optional reduction of $H$. For any $\mathbb{F}$ stopping time $T$, $$
\mathbb{E}[\ind_{\{K_T\leq 0\}}Z_T]=\mathbb{E}[\ind_{\{H_T\leq 0\}}\ind_{\{T<{\tau}\}}]= 0.
$$
By section theorems (cf. \cite{HWY}), $\ind_{\{K\leq 0\}}Z= 0$, proving the first assertion. The second assertion can be proved in a similar way.
\end{proof}

\bl\label{TrU}
For any $\mathbb{G}$ stopping time $U$, there exists an $\mathbb{F}$ stopping time $T$ such that $T\wedge{\tau}=U\wedge{\tau}$.
\el

See \cite[Chapitre XX n$^\circ$75]{DM3} for a proof of the above lemma. We can say more in the case of predictable stopping times.

\bl\label{predictS}
Let $U$ be a $\mathbb{G}$ predictable stopping time. There exists an $\mathbb{F}$ predictable stopping time $T$ such that $U\wedge {\tau}=T\wedge {\tau}$.
\el

\begin{proof}
Let $(U_n)_{n\in\mathbb{N}}$ be a foretelling sequence of $U$ in $\mathbb{G}$ (cf. \cite[Definition 3.26]{HWY}). Let $(T'_n)_{n\in\mathbb{N}}$ be $\mathbb{F}$ stopping times such that $U_n\wedge {\tau}=T'_n\wedge {\tau}$. We can suppose that $(T'_n)_{n\in\mathbb{N}}$ is non decreasing. Let $$
T=(\sup_{i\in\mathbb{N}}T'_i)_{\{\forall k, T'_k<\sup_{i\in\mathbb{N}}T'_i\}}, \ T_n=(T'_n)_{\{T'_n<\sup_{i\in\mathbb{N}}T'_i\}}\wedge n,\ n\in\mathbb{N}.
$$
Then, $T_n<T$ and $\lim_nT_n=T$, i.e. $T$ is an $\mathbb{F}$ predictable stopping time. This stopping time $T$ meets the lemma, because on the set $\{\forall k, T'_k<\sup_{i\in\mathbb{N}}T'_i\}$,$$
T\wedge {\tau}
=
(\sup_{i\in\mathbb{N}}T'_i)\wedge {\tau}
=
\sup_{i\in\mathbb{N}}(T'_i\wedge {\tau})
=
\sup_{i\in\mathbb{N}}(U_i\wedge {\tau})
=
(\sup_{i\in\mathbb{N}}U_i)\wedge {\tau}
=
U\wedge {\tau},
$$
and because, for any $k\in\mathbb{N}$, $$
\{T'_k=\sup_{i\in\mathbb{N}}T'_i\}\cap\{U_k<{\tau}\}
\subset
\{U_k=\sup_{i\in\mathbb{N}}U_i, \sup_{i\in\mathbb{N}}T'_i<{\tau}\}\cap\{U_k<{\tau}\}
=\emptyset,
$$ 
so that, on the set $\{T'_k=\sup_{i\in\mathbb{N}}T'_i\}$, $U> U_k\geq {\tau}$ and 
$
T\wedge {\tau}
=
{\tau}
=
U\wedge {\tau}.
$
\end{proof}

We now look at the path regularity of the reductions.

\bl\label{redreg}
Suppose that $X$ is a $\mathbb{G}$ class($D$) optional c\`adl\`ag process. Let $Y$ be an $\mathbb{F}$ optional reduction of $X$. Then, the process $Y$ is finite and right continuous on $[0,\zeta]$ and $Y$ has finite left limit on $\cup_n[0,\zeta_n]$. $Y_-\ind_{\{Z_->0\}}$ is an $\mathbb{F}$ predictable reduction of $X_-$.
\el

\begin{proof}
Clearly, $Y$ is c\`adl\`ag on $[0,{\tau})$. But, we will prove better. Taking the $\mathbb{F}$ optional projection, we have ${^{o\cdot\mathbb{F}}}(X\ind_{[0,{\tau})})=YZ$. By \cite[Chapitre VI n$^\circ$47 and n$^\circ$50]{DM2} or \cite[exercices 4.13-4.15]{HWY}, ${^{o\cdot\mathbb{F}}}(X\ind_{[0,{\tau})})$ is a c\`adl\`ag process on $\mathbb{R}_+$. This property implies, on the one hand, that the process $Y$ is finite and right continuous on $\{Z>0\}=[0,\zeta)$ and hence on $[0,\zeta]$. On the other hand, $Y$ has finite left limit on $\{Z_->0\}=\{Z_0>0\}\cap(\cup_n[0,\zeta_n])$. Note that $\{Z_0=0\}\cap(\cup_n[0,\zeta_n])=\{Z_0=0\}\cap[0]$ where the process $Y$ has finite left limit by definition. We can now write $X_-\ind_{(0,{\tau}]}=Y_-\ind_{(0,{\tau}]}$, because $(0,{\tau}]\subset \{Z_{-}>0\}$ (cf. Lemma \ref{asupport}, also \cite{yorlemma}), i.e., $Y_-\ind_{\{Z_->0\}}$ is an $\mathbb{F}$ predictable reduction of $X_-$. 
\end{proof}

We introduce the notion of the $\mathbb{F}$ reduction especially to study the $\mathbb{G}$ local martingales $X$ with $\Delta_{\tau}X=0$.

\bl\label{rdm}
Let $X$ be a $\mathbb{G}$ local martingale with $\Delta_{\tau}X=0$ on $\{0<{\tau}<\infty\}$. Let $Y$ be an $\mathbb{F}$ optional reduction of $X$. Then, $Y$ is an $\mathbb{F}$ semimartingale on $\cup_n[0,\zeta_n]$  (cf. \cite[Definition 8.19]{HWY}) such that $YZ+Y_-{_\centerdot}\mathsf{a}$ or equivalently $Z_-{_\centerdot}Y+[Y,Z]$ are $\mathbb{F}$ local martingales on $\cup_n[0,\zeta_n]$. 
\el

\begin{proof}
Assume firstly that $X^{\tau}$ and $(X^{\tau})_-$ are in class($D$) in $\mathbb{G}$. We write$$
\ind_{\{0<{\tau}\}}X^{\tau}=Y\ind_{[0,{\tau})}+Y_{{\tau}-}\ind_{\{0<{\tau}\}}\ind_{[{\tau},\infty)}.
$$
Consider the $\mathbb{F}$ optional projections of the three processes in this identity. ${^{o\cdot\mathbb{F}}}(\ind_{\{0<{\tau}\}}X^{\tau})$ is an $\mathbb{F}$ martingale. ${^{o\cdot\mathbb{F}}}(Y\ind_{[0,{\tau})})=YZ$. According to \cite[Corollary 5.31]{HWY}, $$
{^{o\cdot\mathbb{F}}}(Y_{{\tau}-}\ind_{\{0<{\tau}\}}\ind_{[{\tau},\infty)})
-
(Y_{{\tau}-}\ind_{\{0<{\tau}\}}\ind_{[{\tau},\infty)}){^{p\cdot\mathbb{F}}}
=
{^{o\cdot\mathbb{F}}}(Y_{{\tau}-}\ind_{\{0<{\tau}\}}\ind_{[{\tau},\infty)})
-
Y_-{_\centerdot}\mathsf{a}
$$ 
is an $\mathbb{F}$ martingale, where $Y_-{_\centerdot}\mathsf{a}$ is well-defined because of Lemma \ref{asupport} and \ref{redreg}. We conclude that $
YZ+Y_-{_\centerdot}\mathsf{a}
$
is an $\mathbb{F}$ martingale. Moreover, for every $\zeta_n$, $Y^{\zeta_n}$ is c\`adl\`ag on $\mathbb{R}_+$ (cf. Lemma \ref{redreg}), and, from the identity (with the convention $\frac{0}{0}=0$)
$$
\dcb
(YZ)^{\zeta_n}\frac{1}{Z^{\zeta_n-}}=Y^{\zeta_n}\ind_{[0,\zeta_n)}+(YZ)_{\zeta_n}\frac{1}{Z_{\zeta_n-}}\ind_{[\zeta_n,\infty)},
\dce
$$
noting that $(YZ)^{\zeta_n}$ and $\ind_{\{Z_0>0\}}\frac{1}{Z^{\zeta_n-}}$ are $\mathbb{F}$ semimartingales, we deduce that$$
\dcb
Y^{\zeta_n}
=
Y^{\zeta_n}\ind_{[0,\zeta_n)}+Y_{\zeta_n}\ind_{[\zeta_n,\infty)}
=
(YZ)^{\zeta_n}\frac{1}{Z^{\zeta_n-}}-(YZ)_{\zeta_n}\frac{1}{Z_{\zeta_n-}}\ind_{[\zeta_n,\infty)}+Y_{\zeta_n}\ind_{[\zeta_n,\infty)}
\dce
$$
is an $\mathbb{F}$ semimartingale. Applying the integration by parts formula on $\cup_n[0,\zeta_n]$, we obtain$$
\dcb
YZ+Y_-{_\centerdot}\mathsf{a}
=
Z_-{_\centerdot}Y+Y_-{_\centerdot}Z+[Y,Z]+Y_-{_\centerdot}\mathsf{a}.
\dce
$$
As $Y_-{_\centerdot}Z+Y_-{_\centerdot}\mathsf{a}=Y_-{_\centerdot}{\replacemdot}$ is an $\mathbb{F}$ local martingale on $\cup_n[0,\zeta_n]$, we conclude that $Z_-{_\centerdot}Y+[Y,Z]$ is an $\mathbb{F}$ local martingale on $\cup_n[0,\zeta_n]$.

Consider now the case where $X$ is a general $\mathbb{G}$ local martingale. Let $(T'_n)_{n\in\mathbb{N}}$ be a sequence of $\mathbb{G}$ stopping times, increasing to the infinity, such that every $X^{{\tau}\wedge T'_n}$ and $(X^{{\tau}\wedge T'_n})_-$ are processes in class($D$) in $\mathbb{G}$. Let $a>0$ be a real number and suppose $
\mathbb{Q}[T'_n<a]\leq\frac{1}{2^n},\ n\in\mathbb{N}_+.
$
By Lemma \ref{TrU}, there exists an $\mathbb{F}$ stopping time $T_n$ such that $T_n\wedge{\tau}=T'_n\wedge{\tau}$. Then, $Y^{T_n}$ is an $\mathbb{F}$ optional reduction of $X^{T'_n}$, and for any positive integer $k$, the preceding results implies that $Y^{T_n\wedge\zeta_k}$ is an $\mathbb{F}$ semimartingale on $\mathbb{R}_+$, satisfying the martingale equation of the lemma on $[0,\zeta_k]$. We have the inequalities:$$
\dcb
&&\sum_{n=1}^\infty\mathbb{Q}[Z_{T_n}>\frac{1}{k}, T_n<a]
\leq
k\sum_{n=1}^\infty\mathbb{E}[Z_{T_n}\ind_{\{T_n<a\}}]
=
k\sum_{n=1}^\infty\mathbb{Q}[T_n<{\tau}, T_n<a]\\
&=&
k\sum_{n=1}^\infty\mathbb{Q}[T'_n<{\tau}, T'_n<a]
\leq
k\sum_{n=1}^\infty\frac{1}{2^n}<\infty.
\dce
$$
The Borel-Cantelli lemma implies that, almost surely, there exists a $T_n$ such that either $Z_{T_n}\leq \frac{1}{k}$, i.e. $\zeta_k\leq T_n$, or $T_n\geq a$. We conclude $T_n\geq \zeta_k\wedge a$ and therefore that $Y^{\zeta_k\wedge a}$ is an $\mathbb{F}$ semimartingale (cf. \cite[Theorem 8.10]{HWY}) satisfying the martingale equation of the lemma. Letting $a\uparrow\infty$, we prove the lemma.
\end{proof}

Lemma \ref{rdm} characterizes the $\mathbb{G}$ local martingales $X$ with $\Delta_{\tau}X=0$ by a martingale equation in $\mathbb{F}$. We rewrite the martingale equation in Lemma \ref{rdm} in the form
$$
YZ+Y_-{_\centerdot}\mathsf{a}
=
YZ+(YZ)_-\frac{1}{Z_-}{_\centerdot}\mathsf{a}
=
\mbox{an $\mathbb{F}$ local martingale on $\cup_n[0,\zeta_n]$}.
$$
The following lemma is therefore useful.

\bl\label{yyam}
Let $X$ be an $\mathbb{F}$ semimartingale on $\cup_n[0,\zeta_n]$. Then, $X$ satisfies the martingale equation
$$
X+X_-\frac{1}{Z_-}{_\centerdot}\mathsf{a}
=
\mbox{an $\mathbb{F}$ local martingale on $\cup_n[0,\zeta_n]$},
$$
and $X_{\ddot{\eta}}=0$ on $\{\ddot{\eta}<\infty\}$, if and only if $X$ has the form $X=\mathcal{E}(-\frac{1}{Z_-}{_\centerdot}\mathsf{a})M$ for an $\mathbb{F}$ local martingale $M$ on $\cup_n[0,\zeta_n]$. Moreover, if $M'$ denotes the $\mathbb{F}$ local martingale at the right hand side of the martingale equation, we have $\Delta_{\ddot{\eta}} M'=0$ on $\{0<\ddot{\eta}<\infty\}$. 
\el

\begin{proof}
Note that the process $\mathcal{E}(-\frac{1}{Z_-}{_\centerdot}\mathsf{a})$ is well-defined on $\cup_n[0,\zeta_n]=[0]\cup\{Z_->0\}$. For $s\in\cup_n[0,\zeta_n]$, $Z_{s-}>0$ and $-\frac{1}{Z_{s-}}\Delta_s\mathsf{a}=-1$, if and only if $s=\ddot{\eta}$. We write on $\cup_n[0,\zeta_n]$$$
\mathcal{E}(-\frac{1}{Z_-}{_\centerdot}\mathsf{a})
=
\mathcal{E}(-\ind_{[0,\ddot{\eta})}\frac{1}{Z_-}{_\centerdot}\mathsf{a})
\mathcal{E}(-\ind_{[\ddot{\eta}]}\frac{1}{Z_-}{_\centerdot}\mathsf{a})
=
\mathcal{E}(-\ind_{[0,\ddot{\eta})}\frac{1}{Z_-}{_\centerdot}\mathsf{a})\ind_{[0,\ddot{\eta})}.
$$ 
The stochastic exponential $\mathcal{E}(-\ind_{[0,\ddot{\eta})}\frac{1}{Z_-}{_\centerdot}\mathsf{a})$ is a $\mathbb{F}$ predictable (cf. Lemma \ref{ttC} and \cite[Theorem 3.33]{HWY}) non null non increasing process on $\cup_n[0,\zeta_n]$ so that $\mathcal{E}(-\ind_{[0,\ddot{\eta})}\frac{1}{Z_-}{_\centerdot}\mathsf{a})^{-1}$ is a $\mathbb{F}$ predictable non null non decreasing process on $\cup_n[0,\zeta_n]$.

We check by the integration by parts formula and Yoeurp's lemma that if $X$ is given by $X=\mathcal{E}(-\frac{1}{Z_-}{_\centerdot}\mathsf{a})M$ on $\cup_n[0,\zeta_n]$, the process $X$ satisfies the equation. In this case, the $\mathbb{F}$-martingale at the right hand side of the equation is given by $M_0+\mathcal{E}(-\frac{1}{Z_-}{_\centerdot}\mathsf{a}){_\centerdot}M$. Clearly $X_{\ddot{\eta}}=0$.

Suppose now that $X$ satisfies the martingale equation with an $\mathbb{F}$ martingale $M'$ on $\cup_n[0,\zeta_n]$ at the right hand side of the martingale equation, and $X_{\ddot{\eta}}=0$. Then,$$
\Delta_{\ddot{\eta}} M'= \Delta_{\ddot{\eta}}X+X_{\ddot{\eta}-}=0
$$
on $\{0<\ddot{\eta}<\infty\}$. Set $
M=\mathcal{E}(-\ind_{[0,\ddot{\eta})}\frac{1}{Z_-}{_\centerdot}\mathsf{a})^{-1}{_\centerdot}M'
$
on $\cup_n[0,\zeta_n]$. Then $M$ satisfies $$
M'=M'_0+\mathcal{E}(-\frac{1}{Z_-}{_\centerdot}\mathsf{a}){_\centerdot}M
\ \mbox{ on $\cup_n[0,\zeta_n]$.}
$$ 
Hence, the martingale equation becomes
$$
X+X_-\frac{1}{Z_-}{_\centerdot}\mathsf{a}
=
M'_0+\mathcal{E}(-\frac{1}{Z_-}{_\centerdot}\mathsf{a}){_\centerdot}M.
$$
By the uniqueness of the solution of stochastic differential equation (cf. \cite{protter}), $X$ must be given by $\mathcal{E}(-\frac{1}{Z_-}{_\centerdot}\mathsf{a})(M'_0+M)$ ($M_0=0$) on $\cup_n[0,\zeta_n]$.
\end{proof}

\brem
Notice that the Az\'ema supermartingale $Z$ satisfies the martingale equation of Lemma \ref{yyam} with the local martingale ${\replacemdot}$ at the right hand side. Consequently, $$
Z=\mathcal{E}(-\frac{1}{Z_-}{_\centerdot}\mathsf{a})(Z_0+\mathcal{E}(-\frac{1}{Z_-}{_\centerdot}\mathsf{a})^{-1}{_\centerdot}{\replacemdot}) \ \mbox{ on $\mathtt{C}(\frac{1}{{\replacegamma}})$},
$$ 
that yields another proof of Lemma \ref{newMdecomp}. In fact, it was the original proof. 
\erem

We present the counterparty to Lemma \ref{rdm}.

\bl\label{csinv}
For any $\mathbb{F}$ semimartingale $X$ on $\cup_n[0,\zeta_n]$ such that $XZ+X_-{_\centerdot}\mathsf{a}$ or equivalently $Z_-{_\centerdot}X+[X,Z]$ are $\mathbb{F}$ local martingales on $\cup_n[0,\zeta_n]$, $X^{{\tau}-}$ is a $\mathbb{G}$ local martingale.
\el

\brem
Note that, if $X$ satisfies the condition of the lemma, $$
\Delta_{\zeta}(Z_-{_\centerdot}X+[X,Z])
=
Z_{\zeta-}\Delta_{\zeta}X+\Delta_{\zeta}X\Delta_{\zeta}Z
=0, \ \mbox{ if $0<\zeta<\infty$ and $\zeta\in\cup_n[0,\zeta_n]$}.
$$
This means that $X^{\zeta-}$ also satisfies the condition of the lemma.
\erem

\begin{proof}
Note that $X^{{\tau}-}$ is well defined, because ${\tau}\in\cup_n[0,\zeta_n]$. Let $U$ be an $\mathbb{F}$ stopping time $\leq \zeta_n$ for some $n$, reducing $X$ and $XZ+X_-{_\centerdot}\mathsf{a}$, making $X^U_-$ bounded. We compute$$
\dcb
\mathbb{E}[X^{{\tau}-}_U\ind_{\{0<{\tau}\}}]
&=&
\mathbb{E}[X_{U}\ind_{\{U<{\tau}\}}+X_{{\tau}-}\ind_{\{0<{\tau}\leq U\}}]
=
\mathbb{E}[X_{U}Z_{U}+\int_0^{U}X_{s-}d\mathsf{a}_s]
=
\mathbb{E}[X_0Z_0].
\dce
$$
We prove that $(X^{{\tau}-})^{\zeta_n}$ is a $\mathbb{G}$ local martingale (cf. \cite[Theorem 4.40]{HWY}). As $\{\zeta_n<{\tau}\}\downarrow\emptyset$, the lemma is proved.
\end{proof}

The following lemma will not directly used in this paper. But it is a natural continuation of the preceding results.

\bl\label{rdi}
Let $A$ be a $\mathbb{G}$ predictable process with finite variation null at the origin. Let $B$ be an $\mathbb{F}$ predictable reduction of $A$, null at the origin. Then, on $\cup_n[0,\zeta_n]$, $B$ is a c\`adl\`ag process and has finite variation.
\el

\begin{proof}
We only consider $A$ which is non negative, non decreasing. We note that, as $B_0=0$, $B$ is also an $\mathbb{F}$ optional reduction of $A$. 

We suppose for the moment that $A$ is integrable. Lemma \ref{redreg} is applicable and, consequently, $B$ is a c\`adl\`ag process on $\cup_n[0,\zeta_n]$. For real numbers $0<a\leq b$ we write $$
B_a\ind_{\{b\leq{\tau}\}}=A_a\ind_{\{b\leq{\tau}\}}
\leq
A_b\ind_{\{b\leq{\tau}\}}=B_b\ind_{\{b\leq{\tau}\}}.
$$
Taking conditional expectation with respect to $\mathcal{F}_{b-}$ we obtain $
B_aZ_{b-}
\leq
B_bZ_{b-}
$
and conclude that $B$ is a non decreasing process on $\{Z_0>0\}\cap(\cup_n[0,\zeta_n])$. On $\{Z_0=0\}\cap(\cup_n[0,\zeta_n])=\{Z_0=0\}\cap[0]$, $B=0$ is a non decreasing process by definition.

Consider now the general case. Let $(T'_n)_{n\in\mathbb{N}}$ be a sequence of $\mathbb{G}$ stopping times, increasing to the infinity, such that every $A_{T'_n}$ is integrable. Let $(T_n)_{n\in\mathbb{N}}$ be an increasing sequence of $\mathbb{F}$ stopping times such that $T_n\wedge{\tau}=T'_n\wedge{\tau}$. We check, with $A_0=B_0=0$, that $B^{T_n}$ is an $\mathbb{F}$ predictable reduction of $A^{T'_n}$. The preceding result implies that, for every non negative integer $k$, $(B^{T_n})^{\zeta_k}$ is a non decreasing process. As in the proof of Lemma \ref{rdm}, we fix a real $a>0$ and we can suppose that $\zeta_k\wedge a\leq T_n$ for some $n$. Hence, $B^{\zeta_k\wedge a}$ is a non decreasing process. Letting $a\uparrow\infty$ we prove the lemma.
\end{proof}

\

\section{$\mathbb{G}$ deflators for $\mathbb{F}$ semimartingales stopped at ${\tau}-$}\label{ST-}

We regard the $\mathbb{F}$ semimartingales $X$ and we try to construct deflators for $X^{{\tau}-}$ in $\mathbb{G}$.

\subsection{When $X$ and $X^{\eta-}$ are $\mathbb{F}$ local martingales}

We begin with a particular situation where we compute directly without passing through the reduction mechanism. We make use of this computation to explain how the process $\frac{1}{L^{{\tau}-}}$ plays its role of deflator, before a later general discussion where Lemma \ref{newMdecomp} and Lemma \ref{yyam} will be necessary. Recall the $\mathbb{F}$ local martingale $\mathsf{n}$ defined in Lemma \ref{nmart}.

\bl\label{nXmart}
For any $\mathbb{F}$ optional process $X$ such that $X^{\eta-}$ is an $\mathbb{F}$ local martingale, $M=\mathsf{n}X$ is an $\mathbb{F}$ local martingale.
\el

\begin{proof}
We note that $M=\mathsf{n}X^{\eta-}$. we write with the integration by parts formula$$
\dcb
dM_t
&=&
\mathsf{n}_{t-}dX^{\eta-}_t+X^{\eta-}_{t-}d\mathsf{n}_t+d[\mathsf{n},X^{\eta-}]_t
\dce
$$
We only need to consider the last term. We write $$
\mathsf{n}=\mathcal{E}(-\replacemathsfd)^{-1}-\mathcal{E}(-\replacemathsfd)^{-1}_\eta\ind_{[\eta,\infty)}
$$
because $\replacemathsfd$ is constant on $[\eta,\infty)$. From this we deduce$$
d[\mathsf{n},X^{\eta-}]_t
=
d[\mathcal{E}(-\replacemathsfd),X^{\eta-}]_t
$$
which is an $\mathbb{F}$ local martingale because of Yoeurp's lemma.
\end{proof}

\bl\label{Meta}
For any $\mathbb{F}$ local martingale $M$ such that $M_\eta=0$ on $\{\eta<\infty\}$, $\frac{M^{{\tau}-}}{L^{{\tau}-}}$ is a $\mathbb{G}$ local martingale.
\el

\begin{proof}
Take a finite $\mathbb{F}$ stopping time $\sigma$ localizing $M$ and $D$ such that $\sigma\leq \zeta_n$ for some $n$. Applying Lemma \ref{daLD}, we write$$
\dcb
\mathbb{E}[\frac{M^{{\tau}-}_\sigma}{L^{{\tau}-}_\sigma}\ind_{\{0<{\tau}\}}]
&=&
\mathbb{E}[\frac{M_\sigma}{L_\sigma}\ind_{\{\sigma<{\tau}\}}]
+
\mathbb{E}[\frac{M_{{\tau}-}}{L_{{\tau}-}}\ind_{\{0<{\tau}\leq \sigma\}}]\\

&=&
\mathbb{E}[\frac{M_\sigma}{L_\sigma} Z_\sigma]
-
\mathbb{E}[\int_0^\sigma\frac{M_{s-}}{L_{s-}}\ind_{\{s\in\mathtt{C}(\frac{1}{{\replacegamma}})\cup[\ddot{\eta}]\}}L_{s-}dD_s].
\dce
$$
Note that $Z_\sigma=Z_\sigma\ind_{\{\sigma<\zeta\}}$ so that $\frac{M_\sigma}{L_\sigma} Z_\sigma$ is well-defined. When $\sigma<\zeta$, we have $Z_\sigma=L_\sigma D_\sigma$ and $L_{\sigma}>0$. We can write$$
\dcb
&&\frac{M_\sigma}{L_\sigma} Z_\sigma
=
M_\sigma \ind_{\{\sigma<\zeta\}} D_\sigma
=
M_\sigma D_\sigma\ind_{\{0<\zeta\}} - M_\sigma \ind_{\{0<\zeta\leq \sigma\}} D_\sigma,
\dce
$$
and if $\sigma<\infty$,
$$
\dcb
M_\sigma \ind_{\{0<\zeta\leq \sigma\}} D_\sigma
&=&
M_\sigma \ind_{\{\sigma= \zeta=\zeta_n>0\}} D_\sigma
=
M_\zeta \ind_{\{\sigma= \zeta=\zeta_n>0\}}\ind_{\{Z_{\zeta-}>0, D_\zeta>0\}} D_\zeta\\

&=&
M_\eta \ind_{\{\sigma= \zeta=\zeta_n>0\}}\ind_{\{Z_{\zeta-}>0, D_\zeta>0\}} D_\zeta
=
0	.
\dce
$$
Note that $L_->0$ on $\mathtt{C}(\frac{1}{{\replacegamma}})\cup[\ddot{\eta}]=\cup_n[0,\zeta_n]$ (cf. Lemma \ref{Llimit}) and $\ind_{\{s\in\mathtt{C}(\frac{1}{{\replacegamma}})\cup[\ddot{\eta}]\}}dD_s
=
\ind_{\{s\leq \zeta\}}dD_s$ (with the new definition of $D$). Hence
$$
\dcb
\mathbb{E}[\frac{M^{{\tau}-}_\sigma}{L^{{\tau}-}_\sigma}\ind_{\{0<{\tau}\}}]
&=&
\mathbb{E}[M_\sigma D_\sigma\ind_{\{0<\zeta\}}]
-
\mathbb{E}[\int_0^\sigma M_{s-}\ind_{\{s\leq \zeta\}}dD_s]\\

&=&
\mathbb{E}[M_\sigma D_\sigma\ind_{\{0<\zeta\}}]
-
\mathbb{E}[M_\sigma(D_\sigma-D_0)\ind_{\{0<\zeta\}}]
=
\mathbb{E}[M_0 D_0\ind_{\{0<\zeta\}}].
\dce
$$
Let $(\sigma_n)_{n\in\mathbb{N}}$ be a sequence of $\mathbb{F}$ stopping times localizing $M$ and $D$ tending to the infinity. The above computations imply that the sequence of $
(\sigma_n\wedge \zeta_n)_{\{\sigma_n\wedge \zeta_n<{\tau}\}}
$
is a sequence of $\mathbb{G}$ stopping times localizing $\frac{M^{{\tau}-}}{L^{{\tau}-}}$ tending to the infinity and making it a $\mathbb{G}$ local martingale.
\end{proof}

Note that $\mathsf{n}^{{\tau}-}=\frac{1}{\mathcal{E}(-\replacemathsfd)^{{\tau}-}}>0$ on $\{\tau<\infty\}$. Combining Lemmas \ref{nXmart} and \ref{Meta}, we conclude

\bethe\label{dfet}
For any $\mathbb{F}$ local martingale $X$ such that $X^{\eta-}$ also is an $\mathbb{F}$ local martingale, $\frac{\mathsf{n}^{{\tau}-}}{L^{{\tau}-}}=\frac{1}{\mathcal{E}(-\replacemathsfd)^{{\tau}-}L^{{\tau}-}}$ is a $\mathbb{G}$ deflator of $X^{{\tau}-}$.
\ethe

\subsection{Miscellaneous properties of $\eta$ and of $L$}

\bl\label{nonjump}
If $\mathbb{P}[0<\eta<\infty]>0$, there exists an $\mathbb{F}$ local martingale $M$ such that $M^{{\tau}-}$ has the arbitrage of the first kind.
\el

\begin{proof}
From Lemma \ref{Llimit}, Lemma \ref{expressiongamma} and Lemma \ref{nmart}, on the set $\{\eta<\infty\}$, $L_{\eta-}D_{\eta-}=Z_{\eta-}>0, D_\eta>0$.
Let $M=\ind_{\{0<\eta\}}\ind_{[\eta,\infty)}-\replacemathsfd$ so that $M^{{\tau}-}=-\ind_{[0,{\tau})}{_{\centerdot}}\replacemathsfd$. Applying Lemma \ref{nmart}, we obtain$$
\dcb
\mathbb{E}[M^{{\tau}-}_\infty]
&=&
\mathbb{E}[M_{{\tau}-}]
=
\mathbb{E}[-\replacemathsfd_{{\tau}-}]
=
-\mathbb{E}[\ind_{[0,{\tau})}{_{\centerdot}}\replacemathsfd]
=
-\mathbb{E}[{\replacegamma}_{_{\centerdot}}\replacemathsfd]
=
-\mathbb{E}[{\replacegamma}_\eta\ind_{0<\eta<\infty}]<0=\mathbb{E}[M^{{\tau}-}_0].
\dce
$$
This means that $M^{{\tau}-}$ is a $\mathbb{G}$ optional non constant non increasing process, which can not satisfy $\mathtt{NA}_1$ condition.
\end{proof}

\bl
For a non negative $\mathbb{F}$ local martingale $M$, $\frac{M^{{\tau}-}}{L^{{\tau}-}}$ is a $\mathbb{G}$ supermartingale.
\el

\begin{proof}
We consider two $\mathbb{F}$ stopping times $\sigma\leq \iota$ localizing $M$ and $Z$. Let $B\in\mathcal{F}_{\sigma}$. We compute $$
\dcb
\mathbb{E}[\ind_B\frac{M^{{\tau}-}_{\iota}}{L^{{\tau}-}_{\iota}}\ind_{\{\sigma<{\tau}\}}]
&=&
\mathbb{E}[\ind_B\frac{M_\iota}{L_\iota}\ind_{\{\iota<{\tau}\}}]
+
\mathbb{E}[\ind_B\frac{M_{{\tau}-}}{L_{{\tau}-}}\ind_{\{\sigma<{\tau}\leq \iota\}}]\\

&=&
\mathbb{E}[\ind_B\frac{M_\iota}{L_\iota} Z_\iota]
-
\mathbb{E}[\ind_B\int_\sigma^\iota\frac{M_{s-}}{L_{s-}}\ind_{\{s\leq \zeta\}}L_{s-}dD_s]\\

&=&
\mathbb{E}[\ind_B\frac{M_\iota}{L_\iota} Z_\iota]
-
\mathbb{E}[\ind_B\int_\sigma^\iota M_{s-}\ind_{\{s\leq \zeta\}}dD_s]\ (\mbox{ cf. the proof of Lemma \ref{Meta},})\\

&=&
\mathbb{E}[\ind_B M_\iota D_\iota\ind_{\{\iota<\zeta\}}]
-
\mathbb{E}[\ind_B M_{\iota\wedge\zeta}(D_{\iota\wedge\zeta}-D_{\sigma\wedge\zeta})]\\

&\leq&
\mathbb{E}[\ind_B M_\iota D_\iota\ind_{\{\iota<\zeta\}}]
-
\mathbb{E}[\ind_B M_\iota(D_\iota-D_\sigma)\ind_{\{\iota<\zeta\}}]\\

&=&
\mathbb{E}[\ind_B M_\iota D_\sigma\ind_{\{\iota<\zeta\}}]
\leq
\mathbb{E}[\ind_B M_\iota D_\sigma\ind_{\{\sigma<\zeta\}}]\\

&=&
\mathbb{E}[\ind_B M_\sigma D_\sigma\ind_{\{\sigma<\zeta\}}]
=
\mathbb{E}[\ind_B \frac{M_\sigma}{L_\sigma} Z_\sigma]
=
\mathbb{E}[\ind_B\frac{M^{{\tau}-}_{\sigma}}{L^{{\tau}-}_{\sigma}}\ind_{\{\sigma<{\tau}\}}].
\dce
$$
Let $C\in\mathcal{G}_{\sigma}$. There exists $B\in\mathcal{F}_{\sigma}$ such that $C\cap\{\sigma<{\tau}\}=B\cap\{\sigma<{\tau}\}$. We have
$$
\dcb
\mathbb{E}[\ind_C\frac{M^{{\tau}-}_{\iota}}{L^{{\tau}-}_{\iota}}]

&=&
\mathbb{E}[\ind_B\frac{M^{{\tau}-}_{\iota}}{L^{{\tau}-}_{\iota}}\ind_{\{\sigma<{\tau}\}}]
+
\mathbb{E}[\ind_C\frac{M_{{\tau}-}}{L_{{\tau}-}}\ind_{\{{\tau}\leq \sigma\}}]\\

&\leq&
\mathbb{E}[\ind_B\frac{M^{{\tau}-}_{\sigma}}{L^{{\tau}-}_{\sigma}}\ind_{\{\sigma<{\tau}\}}]
+
\mathbb{E}[\ind_C\frac{M_{{\tau}-}}{L_{{\tau}-}}\ind_{\{{\tau}\leq \sigma\}}]
=
\mathbb{E}[\ind_C\frac{M^{{\tau}-}_{\sigma}}{L^{{\tau}-}_{\sigma}}].
\dce
$$
\end{proof}

\subsection{When $X$ is a general $\mathbb{F}$ local martingale}\label{generalsm}

\bethe\label{twoequations}	
For a multi-dimensional $\mathbb{F}$ local martingale $X$, $X^{{\tau}-}$ has a deflator in $\mathbb{G}$ if and only if the following system of martingale problems have solutions: for $\mathsf{z}$ an $\mathbb{F}$ special semimartingale on $\cup_n[0,\zeta_n]$, 
$$
\left\{
\begin{array}{rll}
\mathsf{z}Z+\mathsf{z}_-{_\centerdot}\mathsf{a}& \mbox{is an $\mathbb{F}$ local martingale on $\cup_n[0,\zeta_n]$},\\

[X,\mathsf{z}Z]
& \mbox{is an $\mathbb{F}$ local martingale on $\cup_n[0,\zeta_n]$},\\

\mathsf{z}>0 &\mbox{on $[0,\zeta)$,}\\

\mathsf{z}_->0 &\mbox{on $\cup_n[0,\zeta_n]$}.
\dce
\right.
$$
In this case, $
\Phi=\mathsf{z}^{{\tau}-}
$
is a deflator for $X^{{\tau}-}$ in $\mathbb{G}$.
\ethe

\begin{proof}
\textit{Necessity} 
Let $\Phi$ be a deflator of $X^{{\tau}-}$ in $\mathbb{G}$. According to Corollary \ref{DY=0} we can suppose $\Delta_{\tau}\Phi=0$ on $\{0<{\tau}<\infty\}$. We suppose also $\Phi_0=1$. Let $\mathsf{z}$ be an $\mathbb{F}$ optional reduction of $\Phi^{{\tau}}=\Phi^{{\tau}-}$. We can suppose $\mathsf{z}_0\equiv 1$. According to Lemma \ref{rdm}, $\mathsf{z}$ is an $\mathbb{F}$ special semimartingale on $\cup_n[0,\zeta_n]$ with $\mathsf{z}Z+\mathsf{z}_-{_\centerdot}\mathsf{a}$ to be an $\mathbb{F}$ local martingale on $\cup_n[0,\zeta_n]$. According to Lemma \ref{redreg}, $\mathsf{z}_-\ind_{\{Z_->0\}}$ is an $\mathbb{F}$ predictable reduction of $\Phi_-$. Applying Lemma \ref{rdps}, $\mathsf{z}>0$ on $[0,\zeta)$ and $\mathsf{z}_->0$ on $\cup_n[0,\zeta_n]\setminus\{0\}$. But, $\mathsf{z}_{0-}=\mathsf{z}_0=1>0$. With the same reasoning, $\mathsf{z}X$ (which is an $\mathbb{F}$ optional reduction of $\Phi^{{\tau}}X^{{\tau}-}$) is an $\mathbb{F}$ special semimartingale on $\cup_n[0,\zeta_n]$ such that $
\mathsf{z}XZ+(\mathsf{z}X)_-{_\centerdot}\mathsf{a}
$
is an $\mathbb{F}$ local martingale on $\cup_n[0,\zeta_n]$. We have
$$
\mathsf{z}XZ+(\mathsf{z}X)_-{_\centerdot}\mathsf{a}
=
X_-{_\centerdot}(\mathsf{z}Z)
+(\mathsf{z}Z)_-{_\centerdot}X+[X,\mathsf{z}Z]+X_-\mathsf{z}_-{_\centerdot}\mathsf{a}
$$
on $\cup_n[0,\zeta_n]$. We conclude that $[X,\mathsf{z}Z]$ is an $\mathbb{F}$ local martingale on $\cup_n[0,\zeta_n]$

\textit{Sufficiency} Suppose that the above martingale problem has a solution $\mathsf{z}$. Set $
\Phi=\mathsf{z}^{{\tau}-}.
$
As $\mathsf{z}$ and $\mathsf{z}X$ satisfy the equation in Lemma \ref{csinv}, $\Phi$ and $\Phi X^{{\tau}-}$ are $\mathbb{G}$ local martingales. This shows that $\Phi$ is a deflator of $X^{{\tau}-}$ in $\mathbb{G}$, because $\Phi>0$. 
\end{proof}

\bethe\label{crochetxm}
For a multi-dimensional $\mathbb{F}$ local martingale $X$, $X^{{\tau}-}$ has a deflator in $\mathbb{G}$ if and only if there exists an $\mathbb{F}$ optional process $M$ such that the pair $(X,M)$ satisfies
\ebe
\item[.]
$M$ is an $\mathbb{F}$ local martingale on $\cup_n[0,\zeta_n]$,
\item[.]
$M>0$ on $[0,\zeta)$, $M_->0$ on $\{Z_0>0\}\cap(\cup_n[0,\zeta_n])$, $M_\eta=0$ on $\{\eta<\infty\}$ and 
\item[.]
$X^{\ddot{\eta}-}M$ is an $\mathbb{F}$ local martingale on $\cup_n[0,\zeta_n]$.
\dbe
Under these conditions, $\ind_{\{Z_0=0\}}+\ind_{\{Z_0>0\}}\frac{1}{Z_0}\frac{M^{{\tau}-}}{L^{{\tau}-}}$ is a deflator for $X^{{\tau}-}$ in $\mathbb{G}$.
\ethe

\brem
Note that $M^{\ddot{\eta}-}\ind_{\cup_n[0,\zeta_n]}+\ind_{\cap_n(\zeta_n,\infty)}$ satisfies also the above conditions. We can therefore modify the process $M$ so that $\{M=0\}=[\eta]$. 
\erem

\begin{proof}
Suppose that $X^{{\tau}-}$ has a deflator in $\mathbb{G}$. Let $\mathsf{z}$ be the process satisfying the equations in Theorem \ref{twoequations} and $\mathsf{z}_0=1$. Then, the process $\mathsf{z}Z$ satisfies the equations in Lemma \ref{yyam}. There exists an $\mathbb{F}$ local martingale $M$ on $\cup_n[0,\zeta_n]$ such that $\mathsf{z}Z=\mathcal{E}(-\frac{1}{Z_-}{_\centerdot}\mathsf{a})M$. Clearly $M>0$ on $[0,\zeta)$ and $M_->0$ on $\{Z_0>0\}\cap(\cup_n[0,\zeta_n])$ and $M_\eta=0$ on $\{\eta<\infty\}$. As $\mathcal{E}(-\frac{1}{Z_-}{_\centerdot}\mathsf{a})$ is predictable with finite variation on $\cup_n[0,\zeta_n]$, with help of Yoeurp's lemma, the second martingale equation in Theorem \ref{twoequations} is the same to say that $\mathcal{E}(-\frac{1}{Z_-}{_\centerdot}\mathsf{a}){_\centerdot}[X,M]$, or equivalently $[X^{\ddot{\eta}-},M]$, or again $X^{\ddot{\eta}-}M$ is an $\mathbb{F}$ local martingale on $\cup_n[0,\zeta_n]$ ($X^{\ddot{\eta}-}$ being an $\mathbb{F}$ local martingale because $\ddot{\eta}$ is predictable (cf. \cite[Examples 9.4]{HWY})).

Conversely, define $$
\mathsf{z}
=\ind_{\{Z_0=0\}}+\ind_{\{Z_0>0\}}\mathcal{E}(-\frac{1}{Z_-}{_\centerdot}\mathsf{a})^{\ddot{\eta}-}(\frac{M}{Z})^{\eta-}
$$
on $\cup_n[0,\zeta_n]$. Then, $\mathsf{z}Z=\ind_{\{Z_0>0\}}\mathcal{E}(-\frac{1}{Z_-}{_\centerdot}\mathsf{a})M$ on $\cup_n[0,\zeta_n]$ (because $\mathcal{E}(-\frac{1}{Z_-}{_\centerdot}\mathsf{a})_{\ddot{\eta}}=0,M_\eta=0$). The equations in Theorem \ref{twoequations} are satisfied by $\mathsf{z}$ for $X$, proving the deflator property in $\mathbb{G}$ for $X^{{\tau}-}$.

When the conditions in the theorem are satisfied, $\Phi=\mathsf{z}^{{\tau}-}$ is a deflator of $X^{{\tau}-}$ in $\mathbb{G}$ (cf. Theorem \ref{twoequations}). With Lemma \ref{newMdecomp}, $\ind_{\{Z_0>0\}}\mathcal{E}(-\frac{1}{Z_-}{_\centerdot}\mathsf{a})=\ind_{\{Z_0>0\}}\frac{1}{Z_0}D$ on $\mathtt{C}(\frac{1}{{\replacegamma}})$ and consequently, $$
\Phi
=
\ind_{\{Z_0=0\}}+\ind_{\{Z_0>0\}}\mathcal{E}(-\frac{1}{Z_-}{_\centerdot}\mathsf{a})^{{\tau}-}(\frac{M}{Z})^{{\tau}-}
=
\ind_{\{Z_0=0\}}+\ind_{\{Z_0>0\}}\frac{1}{Z_0}\frac{M^{{\tau}-}}{L^{{\tau}-}}.
$$ 
\end{proof}

\brem
As an application, if $X$ is an $\mathbb{F}$ local martingale having no jump at $\eta$, $X^{\ddot{\eta}-}$ also is an $\mathbb{F}$ local martingale having no jump at $\eta$. Hence, $X^{\ddot{\eta}-}\mathsf{n}$ is an $\mathbb{F}$ local martingale, according to Lemma \ref{nXmart}. With Theorem \ref{crochetxm}, we conclude that $X^{{\tau}-}$ has a deflator in $\mathbb{G}$, as it was proved in Theorem \ref{dfet}.
\erem

\subsection{When $S$ is an $\mathbb{F}$ semimartingale having an $\mathbb{F}$ deflator}\label{SYM}

\bethe\label{whenS}
Let $S$ be a multi-dimensional $\mathbb{F}$ semimartingale having a deflator in $\mathbb{F}$. Let $Y$ be any deflator of $S$ in $\mathbb{F}$. Then, $S^{{\tau}-}$ has a deflator in $\mathbb{G}$ if and only if there exists an $\mathbb{F}$ optional process $M$ such that the triplet $(S,Y,M)$ satisfies
\ebe
\item[.]
$YM$ is an $\mathbb{F}$ local martingale on $\cup_n[0,\zeta_n]$,
\item[.]
$M>0$ on $[0,\zeta)$, $M_->0$ on $\{Z_0>0\}\cap(\cup_n[0,\zeta_n])$, $M_\eta=0$ on $\{\eta<\infty\}$ and 
\item[.]
$S^{\ddot{\eta}-}YM$ is an $\mathbb{F}$ local martingale on $\cup_n[0,\zeta_n]$.
\dbe
\ethe

\begin{proof}
An increasing sequence of bounded $\mathbb{F}$ stopping times $(T_k)_{k\in\mathbb{N}}$ with $T_0=0$ will be called a $Y$-reducing sequence, if $T_k$ tends to the infinity and $Y^{T_k}$ is an $\mathbb{F}$ uniformly integrable martingale. For a $Y$-reducing sequence, we introduce the probability measures $\mathbb{Q}_k=Y^{T_k}{_\centerdot}\mathbb{Q}$. Note that, by Lemma \ref{chgpas}, $[0,\zeta), \cup_n[0,\zeta_n], \{Z_0>0\}, \eta, \ddot{\eta}$ have the same meaning under the probability measures $\mathbb{Q}$ and $\mathbb{Q}_k$. Then,

\vspace{-17pt}

\ebe
\item[]
$S^{{\tau}-}$ has a deflator in $\mathbb{G}$.

\item[$\Leftrightarrow$]
There exists a $Y$-reducing sequence $(T_k)_{k\in\mathbb{N}}$ such that, for any $k\in\mathbb{N}$, $(S^{{\tau}-})^{T_k}=(S^{T_k})^{{\tau}-}$ has a deflator in $\mathbb{G}$.

\item[$\Leftrightarrow$]
There exists a $Y$-reducing sequence $(T_k)_{k\in\mathbb{N}}$ such that, for any $k\in\mathbb{N}$,
$(S^{{\tau}-})^{T_k}=(S^{T_k})^{{\tau}-}$ has a deflator in $\mathbb{G}$ under $\mathbb{Q}_k$.

\item[$\Leftrightarrow$]
There exists a $Y$-reducing sequence $(T_k)_{k\in\mathbb{N}}$ and
$\mathbb{F}$ optional processes $(M_k)_{k\in\mathbb{N}}$ such that, for every $k\in\mathbb{N}$, the pair $(S^{T_k},M_k)$ satisfy the conditions in Theorem \ref{crochetxm} under $\mathbb{Q}_k$ and $\{M_k=0\}=[\eta]$.

\item[$\Leftrightarrow$]
There exists a $Y$-reducing sequence $(T_k)_{k\in\mathbb{N}}$ and
there exists $\mathbb{F}$ optional processes $(M_k)_{k\in\mathbb{N}}$ such that, for every $k\in\mathbb{N}$, the triplet $(S^{T_k},Y^{T_k},M_k)$ satisfy the conditions in the present theorem under $\mathbb{Q}$ and $\{M_k=0\}=[\eta]$.

\item[$\Leftrightarrow$]
There exists an $\mathbb{F}$ optional process $M$ ($M=\prod_{k=1}^\infty\frac{M_k^{T_k}}{M_k^{T_{k-1}}}\ind_{[\eta]^c}$) such that the triplet $(S,Y,M)$ satisfy the conditions in the present theorem.
\dbe
\end{proof}

\bcor
Let $S$ be a multi-dimensional $\mathbb{F}$ semimartingale having a deflator in $\mathbb{F}$. Suppose $\Delta_\eta S=0$ on $\{\eta<\infty\}$. Then, $S^{{\tau}-}$ has a deflator in $\mathbb{G}$
\ecor

\begin{proof}
According to Corollary \ref{DY=0}, we have a deflator $Y$ for $S$ in $\mathbb{F}$ with $\Delta_\eta Y=0$ on $\{\eta<\infty\}$. Let $\mathsf{n}$ be the $\mathbb{F}$ local martingale introduced in Lemma \ref{nmart}. Then, we check the conditions in Theorem \ref{whenS} for the triplet $(S,Y,M=\mathsf{n})$ with help of Lemma \ref{nXmart}.
\end{proof}

\

\section{$\mathbb{G}$ Deflators for $\mathbb{F}$ semimartingales stopped at ${\tau}$}\label{ST}

In this section we reconsider the main Theorem 1.2 of \cite{AFK}. There exists an early note \cite{songnote} on the question, where $Z$ is supposed to be strictly positive. The studies in section \ref{ZP} and section \ref{JRO} enable us now to generalize the approach of that note to give a different proof of Theorem 1.2 of \cite{AFK}.

\subsection{Continuous deflator}

We look for deflators for $X^{{\tau}}$ in $\mathbb{G}$, when $X$ is an $\mathbb{F}$ continuous local martingale. Recall that $\widetilde{Z}$ is the $\mathbb{F}$ optional projection of $\ind_{[0,{\tau}]}$. We have $\widetilde{Z}_0=1$ and $
\widetilde{Z}=Z+\Delta \mathsf{A}
=Z_- +\Delta {\replacem}
$ 
on $(0,\infty)$, where $\mathsf{A}$ is the $\mathbb{F}$ optional dual projection of $\ind_{\{0<{\tau}\}}\ind_{[{\tau},\infty)}$ and ${\replacem}=Z+\mathsf{A}$ which is an $\mathbb{F}$ $\mathtt{BMO}$ martingale.

\bethe\label{continuousdeflator}
Let ${\replaceMtildec}$ denote the $\mathbb{G}$ continuous martingale part of ${\replacem}^{{\tau}}$. Then, the process $-\ind_{[0,{\tau}]}\ind_{\{Z_0>0\}}\frac{1}{Z_-}$
is ${\replaceMtildec}$ integrable in $\mathbb{G}$ and $\mathcal{E}(-\ind_{[0,{\tau}]}\ind_{\{Z_0>0\}}\frac{1}{Z_-}{_\centerdot}{\replaceMtildec})$ is a $\mathbb{G}$ deflator of $X^{{\tau}}$ for any $\mathbb{F}$ continuous local martingale $X$.
\ethe

\begin{proof}
The integrability of $\ind_{[0,{\tau}]}\ind_{\{Z_0>0\}}\frac{1}{Z_-}$ is because it is a bounded on every $[0,\zeta_n]$ for any positive integer $k$, and because $\{\zeta_n<{\tau}\}\downarrow\emptyset$. We need to check that $X^{{\tau}}-[\ind_{[0,{\tau}]}\ind_{\{Z_0>0\}}\frac{1}{Z_-}{_\centerdot}{\replaceMtildec},X^{{\tau}}]$ is a $\mathbb{G}$ local martingale. But the latter is true, because of the relation:$$
\dcb
X^{{\tau}}-[\ind_{[0,{\tau}]}\ind_{\{Z_0>0\}}\frac{1}{Z_-}{_\centerdot}{\replaceMtildec},X^{{\tau}}]
&=&
X^{{\tau}}-\ind_{[0,{\tau}]}\ind_{\{Z_0>0\}}\frac{1}{Z_-}{_\centerdot}\cro{{\replaceMtildec},X}
=
X^{{\tau}}-\ind_{[0,{\tau}]}\frac{1}{Z_-}{_\centerdot}\cro{{\replacem},X}
\dce
$$ 
and of the filtration enlargement formula in \cite[Chapitre XX n$^\circ$76]{DM3} (together with \cite[Chapitre XX n$^\circ$12 and n$^\circ$75]{DM3}) (noting that $\ind_{\{Z_0=0\}}{\replacem}=0$ and the predictable bracket $\cro{{\replacem},X}$ does not depend on the filtrations thanks to the continuity).
\end{proof}

\

\subsection{Purely discontinuous deflator}

We continue the study with purely discontinuous $\mathbb{F}$ local martingales. 

\bl\label{predictkey}
Let $T$ be an $\mathbb{F}$ predictable stopping time. Let $\xi$ be an integrable $\mathcal{F}_T$ measurable random variable. Then,$$
\mathbb{E}[\xi|\mathcal{G}_{T-}]\ind_{\{0<{\tau}, T\leq {\tau}\}}
=
\xi\ind_{\{0=T< {\tau}\}}+\frac{\mathbb{E}[\xi \widetilde{Z}_T|\mathcal{F}_{T-}]}{Z_{T-}}\ind_{\{0<T\leq {\tau}\}}.
$$

\begin{proof}
For any $\mathbb{F}$ stopping time $T'$, $$
\mathbb{E}[\xi|\mathcal{G}_{T'}]\ind_{\{T'< {\tau}\}}
=
\frac{\mathbb{E}[\xi \ind_{\{T'< {\tau}\}}|\mathcal{F}_{T'}]}{Z_{T'}}\ind_{\{T'< {\tau}\}},
$$
known as the key lemma (cf. \cite{BJR} or \cite[Chapitre XX n$^\circ$75]{DM3}). As $T$ is predictable, we can let $T'$ foretelling $T$ to obtain (cf. \cite[Theorem 3.4 and 4.34]{HWY}, \cite[Corollary 2.4]{RY})
$$
\dcb
\mathbb{E}[\xi|\mathcal{G}_{T-}](\ind_{\{0=T< {\tau}\}}+\ind_{\{0<T\leq {\tau}\}})
&=&
\frac{\mathbb{E}[\xi \ind_{\{0< {\tau}\}}|\mathcal{F}_{0}]}{Z_{0}}\ind_{\{0=T< {\tau}\}}+\frac{\mathbb{E}[\xi \ind_{\{T\leq {\tau}\}}|\mathcal{F}_{T-}]}{Z_{T-}}\ind_{\{0<T\leq {\tau}\}}\\\

&=&
\xi\ind_{\{0=T< {\tau}\}}+\frac{\mathbb{E}[\xi \widetilde{Z}_T|\mathcal{F}_{T-}]}{Z_{T-}}\ind_{\{0<T\leq {\tau}\}}.
\dce
$$
Notice that, by Lemma \ref{Llimit}, $Z_{T-}>0$ on $\{0<T\leq {\tau}\}$.
\end{proof}
\el

\bl\label{puredisc}
For any $\mathbb{F}$ purely discontinuous local martingales $X$, the $\mathbb{G}$ martingales part of $X^{\tau}$ also is purely discontinuous.
\el

\begin{proof}
The result is clear if $X$ has finite variation. If $X$ is an $\mathbb{F}$ purely discontinuous square integrable martingale. According to \cite[Theorem 6.22]{HWY}, there exists a sequence of $\mathbb{F}$ martingales $(X_n)_{n\in\mathbb{N}}$ of finite variations such that $\mathbb{E}[[X-X_n,X-X_n]_t]$ tends to zero for any $t\in\mathbb{R}_+$. By \cite[Corollaire(1.8)]{Jeulin80}, the $\mathbb{G}$ martingale part of $X_n$ converges to the $\mathbb{G}$ martingale part of $X$ in the space of square integrable martingales.
\end{proof}

We now introduce 
\ebe
\item[.]
the $\mathbb{F}$ stopping time $\tilde{\eta}=\zeta_{\{0<\zeta<\infty,\widetilde{Z}_\zeta=0<Z_{\zeta-}\}}$, 
\item[.]
the process $\tilde{\replacemathsfd}$ denoting the $\mathbb{F}$ predictable dual projection of $\ind_{\{0<\tilde{\eta}\}}\ind_{[\tilde{\eta},\infty)}$, and 
\item[.]
the process $\tilde{\mathsf{n}}=\mathcal{E}(-\tilde{\replacemathsfd})^{-1}\ind_{[0,\tilde{\eta})}$. 
\dbe
By \cite[Chapitre XX n$^\circ$15]{DM3}, $\widetilde{Z}>0$ on $[0,{\tau}]$. As $\widetilde{Z}_{\tilde{\eta}}=0$ if $\tilde{\eta}<\infty$, necessarily ${\tau}<\tilde{\eta}$. Recall that, by \cite[Lemma 3.5]{AFK}, $\tilde{\mathsf{n}}$ is a well defined $\mathbb{F}$ local martingale with $\{\tilde{\mathsf{n}}>0\}=[0,\tilde{\eta})$ and $\{\tilde{\mathsf{n}}_->0\}=[0,\tilde{\eta}]$. It was a key element in the deflator construction for $X^{{\tau}}$ in \cite{AFK}. Here we use it to introduce the process $$
(\frac{Z_-}{\tilde{\mathsf{n}}_-}\frac{\tilde{\mathsf{n}}}{\widetilde{Z}}
-
1)\ind_{(0,{\tau}]}.
$$ This process has different expressions.
$$
\dcb
&&
(\frac{Z_-}{\tilde{\mathsf{n}}_-}\frac{\tilde{\mathsf{n}}}{\widetilde{Z}}
-
1)\ind_{(0,{\tau}]}

=
(\frac{Z_-}{\tilde{\mathsf{n}}_-}\frac{\tilde{\mathsf{n}}}{\widetilde{Z}}
-
\frac{Z_-}{\tilde{\mathsf{n}}_-}\frac{\tilde{\mathsf{n}}_-}{\widetilde{Z}}
-
\frac{\widetilde{Z}}{\widetilde{Z}}
+
\frac{Z_-}{\widetilde{Z}})\ind_{(0,{\tau}]}\\

&=&

(\frac{Z_-}{\tilde{\mathsf{n}}_-}\frac{\Delta\tilde{\mathsf{n}}}{\widetilde{Z}}
-
\frac{\Delta {\replacem}}{\widetilde{Z}})\ind_{(0,{\tau}]}

=
(\frac{Z_-}{\tilde{\mathsf{n}}_-}\frac{\Delta\tilde{\mathsf{n}}}{\widetilde{Z}}
-
\frac{\Delta {\replacem}}{\widetilde{Z}})\ind_{\{\tilde{\mathsf{n}}_->0\}}\ind_{(0,{\tau}]}.
\dce
$$

\bl
We have ${^{p\cdot\mathbb{G}}}\!(\frac{Z_-}{\tilde{\mathsf{n}}_-}\frac{\tilde{\mathsf{n}}}{\widetilde{Z}}
-
1)\ind_{(0,{\tau}]}=0$.
\el

\begin{proof}
By Lemma \ref{predictS}, we only need to consider $\mathbb{F}$ predictable stopping time $T$ in the computations below. Note that $\tilde{\mathsf{n}}$ is $\mathbb{F}$ locally bounded and $\frac{1}{\tilde{\mathsf{n}}_-}$ is non increasing on the set $\{\tilde{\mathsf{n}}_->0\}$. These facts insure the integrability in the computations below. For any bounded $\mathbb{F}$ stopping time $U$ such that $\tilde{\mathsf{n}}^U$ is a bounded process, applying Lemma \ref{predictkey}, we obtain
$$
\dcb
&&\mathbb{E}[(\frac{Z_{T-}}{\tilde{\mathsf{n}}_{T-}}\frac{\tilde{\mathsf{n}}}{\widetilde{Z}_T}
-
1)\ind_{\{0<T\leq {\tau}\}}|\mathcal{G}_{T-}]\ind_{\{T\leq U\}}\\

&=&
\frac{1}{Z_{T-}}\mathbb{E}[(\ind_{\{\tilde{\mathsf{n}}_{T-}>0\}}\frac{Z_{T-}}{\tilde{\mathsf{n}}_{T-}}\frac{\Delta_T\tilde{\mathsf{n}}}{\widetilde{Z}_T}
-
\ind_{\{\tilde{\mathsf{n}}_{T-}>0\}}\frac{\Delta_T {\replacem}}{\widetilde{Z}_T})\ind_{\{T\leq U\}}\widetilde{Z}_T|\mathcal{F}_{T-}]\ind_{\{0<T\leq {\tau}\}}\\

&=&
\frac{1}{Z_{T-}}\mathbb{E}[(\ind_{\{\tilde{\mathsf{n}}_{T-}>0\}}\frac{Z_{T-}}{\tilde{\mathsf{n}}_{T-}}\Delta_T\tilde{\mathsf{n}}
-
\ind_{\{\tilde{\mathsf{n}}_{T-}>0\}}\Delta_T {\replacem})\ind_{\{T\leq U\}}\ind_{\{\widetilde{Z}_T>0\}}|\mathcal{F}_{T-}]\ind_{\{0<T\leq {\tau}\}}\\

&=&
\frac{1}{Z_{T-}}\mathbb{E}[(\ind_{\{\tilde{\mathsf{n}}_{T-}>0\}}\frac{Z_{T-}}{\tilde{\mathsf{n}}_{T-}}\Delta_T\tilde{\mathsf{n}}
-
\ind_{\{\tilde{\mathsf{n}}_{T-}>0\}}\Delta_T {\replacem})\ind_{\{T\leq U\}}|\mathcal{F}_{T-}]\ind_{\{0<T\leq {\tau}\}}\\

&&
-\frac{1}{Z_{T-}}\mathbb{E}[(\ind_{\{\tilde{\mathsf{n}}_{T-}>0\}}\frac{Z_{T-}}{\tilde{\mathsf{n}}_{T-}}\Delta_T\tilde{\mathsf{n}}
-
\ind_{\{\tilde{\mathsf{n}}_{T-}>0\}}\Delta_T {\replacem})\ind_{\{T\leq U\}}\ind_{\{\widetilde{Z}_T=0<Z_{T-}\}}|\mathcal{F}_{T-}]\ind_{\{0<T\leq {\tau}\}}\\
&&
-
\frac{1}{Z_{T-}}\mathbb{E}[(\ind_{\{\tilde{\mathsf{n}}_{T-}>0\}}\frac{Z_{T-}}{\tilde{\mathsf{n}}_{T-}}\Delta_T\tilde{\mathsf{n}}
-
\ind_{\{\tilde{\mathsf{n}}_{T-}>0\}}\Delta_T {\replacem})\ind_{\{T\leq U\}}\ind_{\{\widetilde{Z}_T=0=Z_{T-}\}}|\mathcal{F}_{T-}]\ind_{\{0<T\leq {\tau}\}}\\

&=&
0\ \mbox{because $\tilde{\mathsf{n}}^U$ and ${\replacem}^U$ are $\mathbb{F}$ uniformly integrable martingales (cf. \cite[Theorem 4.41]{HWY}),}\\

&&

-\frac{1}{Z_{T-}}\mathbb{E}[(\ind_{\{\tilde{\mathsf{n}}_{T-}>0\}}\frac{Z_{T-}}{\tilde{\mathsf{n}}_{T-}}(-\tilde{\mathsf{n}}_{T-})
-
\ind_{\{\tilde{\mathsf{n}}_{T-}>0\}}(-Z_{T-}))\ind_{\{T\leq U\}}\ind_{\{\widetilde{Z}_T=0<Z_{T-}\}}|\mathcal{F}_{T-}]\ind_{\{0<T\leq {\tau}\}}\\
&&
\mbox{because $0<T<\infty, \widetilde{Z}_T=0$ implies $(-Z_{T-})=\Delta_T{\replacem}$,}\\

&&
-
\frac{1}{Z_{T-}}\mathbb{E}[(
-
\ind_{\{\tilde{\mathsf{n}}_{T-}>0\}}\Delta_T {\replacem})\ind_{\{T\leq U\}}\ind_{\{\widetilde{Z}_T=0=Z_{T-}\}}|\mathcal{F}_{T-}]\ind_{\{0<T\leq {\tau}\}}\\

&=&
0
-
0
-
0,
\
\mbox{because $0<T<\infty, \widetilde{Z}_T=0=Z_{T-}$ implies $\Delta_T{\replacem}=0$,}\\
&=&0.
\dce
$$
\end{proof}

\bethe\label{existence2}
There exists a $\mathbb{G}$ purely discontinuous local martingale ${\replacemathsfm}$ such that $$
\Delta{\replacemathsfm}
=
(\frac{Z_-}{\tilde{\mathsf{n}}_-}\frac{\tilde{\mathsf{n}}}{\widetilde{Z}}
-
1)\ind_{(0,{\tau}]}.
$$
\ethe

\begin{proof}
We use the expression
$$
(\frac{Z_-}{\tilde{\mathsf{n}}_-}\frac{\tilde{\mathsf{n}}}{\widetilde{Z}}
-
1)\ind_{(0,{\tau}]}
=
(\frac{Z_-}{\tilde{\mathsf{n}}_-}\frac{\Delta\tilde{\mathsf{n}}}{\widetilde{Z}}
-
\frac{\Delta {\replacem}}{\widetilde{Z}})\ind_{(0,{\tau}]}.
$$
By \cite[Theorem 7.42]{HWY} we need only to check that the non decreasing process$$
\sqrt{\sum_{0<s\leq t\wedge {\tau}}(\frac{Z_-}{\tilde{\mathsf{n}}_-}\frac{\Delta\tilde{\mathsf{n}}}{\widetilde{Z}}
-
\frac{\Delta {\replacem}}{\widetilde{Z}})^2},\ t\in\mathbb{R}_+,
$$
is $\mathbb{G}$ locally integrable. Consider the inequality:
$$
\dcb
&&
\sum_{0<s\leq t\wedge {\tau}}(\frac{\Delta_s {\replacem}}{\widetilde{Z}_s})^2
\leq
\int_0^{t\wedge {\tau}}\frac{d[{\replacem},{\replacem}]}{\widetilde{Z}^2_s}< \infty.
\dce
$$
Let $V_t=\int_0^{t\wedge{\tau}}\frac{d[{\replacem},{\replacem}]}{\widetilde{Z}^2_s}, t\in\mathbb{R}_+$. For $n\in\mathbb{N}_+$, 
we compute
$$
\dcb
&&\mathbb{E}[\Delta_{\zeta_n\wedge {\tau}}\sqrt{V}]
\leq
\mathbb{E}[\frac{|\Delta_{\zeta_n\wedge {\tau}} {\replacem}|}{\widetilde{Z}_{\zeta_n\wedge {\tau}}}]\\
&=&
\mathbb{E}[\frac{|\Delta_{\zeta_n} {\replacem}|}{\widetilde{Z}_{\zeta_n}}\ind_{\{0<\zeta_n\leq {\tau}\}}]
+
\mathbb{E}[\frac{|\Delta_{{\tau}} {\replacem}|}{\widetilde{Z}_{ {\tau}}}\ind_{\{0<{\tau}<\zeta_n\}}]
\leq
\mathbb{E}[|\Delta_{\zeta_n} {\replacem}|]+n\mathbb{E}[|\Delta_{{\tau}} {\replacem}|]
<\infty,
\dce
$$
because ${\replacem}$ is a $\mathtt{BMO}$ martingale.
We conclude that $\sqrt{\sum_{0<s\leq t\wedge {\tau}}(\frac{\Delta_s {\replacem}}{\widetilde{Z}_s})^2}, t\in\mathbb{R}_+$, is $\mathbb{G}$ locally integrable, because $[0,{\tau}]\subset\cup_n[0,\zeta_n]$. 

Similarly, we prove that $$
\sqrt{\sum_{0<s\leq t\wedge {\tau}}(\frac{Z_-}{\tilde{\mathsf{n}}_-}\frac{\Delta\tilde{\mathsf{n}}}{\widetilde{Z}}
)^2}
=
\sqrt{\sum_{0<s\leq t\wedge {\tau}}(\mathcal{E}(-\tilde{\replacemathsfd})_-Z_-\frac{\Delta\tilde{\mathsf{n}}}{\widetilde{Z}}
)^2},\ t\in\mathbb{R}_+,
$$
is $\mathbb{G}$ locally integrable. 
\end{proof}

\bl
$(\frac{Z_-}{\tilde{\mathsf{n}}_-}\frac{\tilde{\mathsf{n}}}{\widetilde{Z}}
-
1)\ind_{(0,{\tau}]}>-1$. The stochastic exponential $\mathcal{E}({\replacemathsfm})$ is strictly positive.
\el

Now we show why this discussion on the process $(\frac{Z_-}{\tilde{\mathsf{n}}_-}\frac{\tilde{\mathsf{n}}}{\widetilde{Z}}
-
1)\ind_{(0,{\tau}]}$.

\bethe\label{mdist2}
Let $M$ be an $\mathbb{F}$ purely discontinuous local martingale such that $\Delta_{\tilde{\eta}} M=0$ on $\{0<\tilde{\eta}<\infty\}$. Then, $\mathcal{E}({\replacemathsfm})M^{\tau}$ is a $\mathbb{G}$ local martingale.
\ethe

\begin{proof}
It is enough to consider the martingale property of $M^{\tau}+[{\replacemathsfm},M^{\tau}]$ in $\mathbb{G}$. Note that $[{\replacem},M]$ and $[\tilde{\mathsf{n}},M]$ have $\mathbb{F}$ locally integrable total variations. For $[{\replacem},M]$, it is the consequence of the $\mathtt{BMO}$ property of ${\replacem}$ and of the Fefferman's inequality (cf. \cite[Theorem 10.18]{HWY}). For $[\tilde{\mathsf{n}},M]$, we need only to note that $\Delta\tilde{\mathsf{n}}$ is $\mathbb{F}$ locally bounded and $M^{\ast}$ is $\mathbb{F}$ locally integrable. Let $T$ be an $\mathbb{F}$ stopping time such that $[{\replacem},M]^T$ and $[\tilde{\mathsf{n}},M]^T$ have integrable total variations, $M^T$ is in class($D$) and $\ind_{\{Z_0>0\}}\frac{1}{Z^T_-}$ is bounded. Note that the jumps of $[{\replacemathsfm},M]$ on $(0,{\tau}]$ are given by$$
\Delta{\replacemathsfm}\Delta M=(\frac{Z_-}{\tilde{\mathsf{n}}_-}\frac{\Delta\tilde{\mathsf{n}}}{\widetilde{Z}}
-
\frac{\Delta {\replacem}}{\widetilde{Z}})\Delta M
=
\frac{Z_{-}}{\tilde{\mathsf{n}}_{-}}\frac{1}{\widetilde{Z}}\Delta[\tilde{\mathsf{n}},M]
-\frac{1}{\widetilde{Z}_s}\Delta[{\replacem},M].
$$ 
We now compute. On the one hand, 
$$
\dcb
&&\mathbb{E}[[{\replacemathsfm},M^{\tau}-M_0]_T]
=
\mathbb{E}[\int_0^{T\wedge{\tau}}d[{\replacemathsfm},M]_s]\\
&=&
\mathbb{E}[\int_0^{T\wedge{\tau}}d[{\replacemathsfm},M^{\tilde{\eta}}]_s]\ 
\mbox{ because ${\tau}<{\tilde{\eta}}$,}\\
&=&
\mathbb{E}[\int_0^{T\wedge {\tau}}\frac{Z_{s-}}{\tilde{\mathsf{n}}_{s-}}\frac{1}{\widetilde{Z}_s}d[\tilde{\mathsf{n}},M^{\tilde{\eta}}]_s]
-\mathbb{E}[\int_0^{T\wedge {\tau}}\frac{1}{\widetilde{Z}_s}d[{\replacem},M^{\tilde{\eta}}]_s]\\

&=&
\mathbb{E}[\int_0^{T}\widetilde{Z}_s\frac{Z_{s-}}{\tilde{\mathsf{n}}_{s-}}\frac{1}{\widetilde{Z}_s}d[\tilde{\mathsf{n}},M^{\tilde{\eta}}]_s]
-\mathbb{E}[\int_0^{T}\widetilde{Z}_s\frac{1}{\widetilde{Z}_s}d[{\replacem},M^{\tilde{\eta}}]_s]\\

&=&
\mathbb{E}[\int_0^{T}\frac{Z_{s-}}{\tilde{\mathsf{n}}_{s-}}\ind_{\{\widetilde{Z}_s>0\}}d[\mathcal{E}(-\tilde{\replacemathsfd})^{-1},M^{\tilde{\eta}}]_s]
-\mathbb{E}[\int_0^{T}\ind_{\{\widetilde{Z}_s>0\}}d[{\replacem},M^{\tilde{\eta}}]_s]\\
&&\mbox{ because $\Delta M^{\tilde{\eta}}=0$ on $[{\tilde{\eta}},\infty)$.}
\dce
$$
For the first term in the last line,
$$
\dcb
&&
\mathbb{E}[\int_0^{T}\frac{Z_{s-}}{\tilde{\mathsf{n}}_{s-}}\ind_{\{\widetilde{Z}_s>0\}}d[\mathcal{E}(-\tilde{\replacemathsfd})^{-1},M^{\tilde{\eta}}]]\\

&=&
\mathbb{E}[\int_0^{T}\mathcal{E}(-\tilde{\replacemathsfd})_{s-}Z_{s-}d[\mathcal{E}(-\tilde{\replacemathsfd})^{-1},M^{\tilde{\eta}}]]
-\mathbb{E}[\int_0^{T}\mathcal{E}(-\tilde{\replacemathsfd})_{s-}Z_{s-}\ind_{\{\widetilde{Z}_s=0\}}d[\mathcal{E}(-\tilde{\replacemathsfd})^{-1},M^{\tilde{\eta}}]]\\

&=&
0\ \mbox{ because of Yoeurp's lemma,}\\
&&-\mathbb{E}[\int_0^{T}\mathcal{E}(-\tilde{\replacemathsfd})_{s-}Z_{s-}\ind_{\{\widetilde{Z}_s=0<Z_{s-}\}}d[\mathcal{E}(-\tilde{\replacemathsfd})^{-1},M^{\tilde{\eta}}]_s]
\\

&=&
0-0\
\mbox{ because $\widetilde{Z}_s=0<Z_{s-}$ implies $s={\tilde{\eta}}$ while $\Delta_{\tilde{\eta}} M=0$.}
\dce
$$
For the second term,
$$
\dcb
&&
-\mathbb{E}[\int_0^{T}\ind_{\{\widetilde{Z}_s>0\}}d[{\replacem},M^{\tilde{\eta}}]_s]\\

&=&

-\mathbb{E}[\int_0^{T}d[{\replacem},M^{\tilde{\eta}}]_s]
+\mathbb{E}[\int_0^{T}\ind_{\{\widetilde{Z}_s=0\}}d[{\replacem},M^{\tilde{\eta}}]_s]
\\

&=&
-\mathbb{E}[\int_0^{T}d[{\replacem},M^{\tilde{\eta}}]_s]
+\mathbb{E}[\int_0^{T}\ind_{\{\widetilde{Z}_s=0<Z_{s-}\}}d[{\replacem},M^{\tilde{\eta}}]_s]
\
\mbox{ because $\ind_{\{\widetilde{Z}=0=Z_{-}\}}\Delta {\replacem}=0$,}\\

&=&
-\mathbb{E}[\int_0^{T}d[{\replacem},M^{\tilde{\eta}}]_s]\
\mbox{ because $\Delta_{\tilde{\eta}} M=0$.}
\dce
$$
On the other hand, by the filtration enlargement formula in \cite[Chapitre XX n$^\circ$76]{DM3},
$$
\dcb
\mathbb{E}[M^{\tau}_T-M_0]
&=&
\mathbb{E}[\int_0^{T\wedge {\tau}}\frac{1}{Z_{s-}}d\cro{{\replacem},M^{\tilde{\eta}}}^{p\cdot\mathbb{F}}_s]
=
\mathbb{E}[\int_0^{T}d\cro{{\replacem},M^{\tilde{\eta}}}^{p\cdot\mathbb{F}}_s]
=
\mathbb{E}[\int_0^{T}d[{\replacem},M^{\tilde{\eta}}]_s]
\dce
$$
(where, as indicated in \cite[Remarque (4.5)]{Jeulin80}, the first and the second equality use implicitly \cite[Lemme (4.3)]{Jeulin80} or Lemma \ref{asupport}). These computations together with Lemma \ref{TrU} and \cite[Theorem 4.40]{HWY} shows that $M^{\tau}+[{\replacemathsfm},M^{\tau}]$ is a $\mathbb{G}$ local martingale.
\end{proof}

\

\subsection{Deflators for $S^{{\tau}}$}

\bethe\label{r++}
For any multi-dimensional $\mathbb{F}$ local martingale $M$ such that $\Delta_{\tilde{\eta}} M=0$ on $\{0<{\tilde{\eta}}<\infty\}$, $M^{{\tau}}$ has a deflator $\mathcal{E}(-\ind_{[0,{\tau}]}\frac{1}{Z_-}{_\centerdot}{\replaceMtildec})\mathcal{E}({\replacemathsfm})$ in $\mathbb{G}$.
\ethe

\begin{proof}
Applying Theorems \ref{continuousdeflator} and \ref{mdist2} component by component, and separately to the continuous part and the purely discontinuous part of $M=M^c+M^d$ in $\mathbb{F}$, we conclude that
$$
M^{{\tau}}\mathcal{E}(-\ind_{[0,{\tau}]}\ind_{\{Z_0>0\}}\frac{1}{Z_-}{_\centerdot}{\replaceMtildec})\mathcal{E}({\replacemathsfm})
=
((M^c)^{{\tau}}+(M^d)^{{\tau}})\mathcal{E}(-\ind_{[0,{\tau}]}\ind_{\{Z_0>0\}}\frac{1}{Z_-}{_\centerdot}{\replaceMtildec})\mathcal{E}({\replacemathsfm})
$$
is a $\mathbb{G}$ local martingale. Consequently, $\mathcal{E}(-\ind_{[0,{\tau}]}\frac{1}{Z_-}{_\centerdot}{\replaceMtildec})\mathcal{E}({\replacemathsfm})$ is a $\mathbb{G}$ deflator for $M^{{\tau}}$.
\end{proof}

Here is a new proof of \cite[Theorem 1.2]{AFK}.

\bcor\label{newafk}
For any multi-dimensional $\mathbb{F}$ semimartingale $S$ having a deflator in $\mathbb{F}$, if $\Delta_{\tilde{\eta}} S=0$ on $\{0<{\tilde{\eta}}<\infty\}$, $S^{{\tau}}$ has a deflator in $\mathbb{G}$. 
\ecor

\begin{proof}
According to Corollary \ref{DY=0}, there exists an $\mathbb{F}$ deflator $Y$ of $S^{{\tilde{\eta}}}$ in $\mathbb{F}$ which has no jump at ${\tilde{\eta}}$. Applying Theorems \ref{r++} to $(Y,SY)$, we conclude that
$$
\dcb
Y^{{\tau}}\mathcal{E}(-\ind_{[0,{\tau}]}\ind_{\{Z_0>0\}}\frac{1}{Z_-}{_\centerdot}{\replaceMtildec})\mathcal{E}({\replacemathsfm})\  \mbox{ and }\

S^{{\tau}}Y^{{\tau}}\mathcal{E}(-\ind_{[0,{\tau}]}\ind_{\{Z_0>0\}}\frac{1}{Z_-}{_\centerdot}{\replaceMtildec})\mathcal{E}({\replacemathsfm})
\dce
$$
are $\mathbb{G}$ local martingales. Consequently, $Y^{{\tau}}\mathcal{E}(-\ind_{[0,{\tau}]}\frac{1}{Z_-}{_\centerdot}{\replaceMtildec})\mathcal{E}({\replacemathsfm})$ is a $\mathbb{G}$ deflator for $S^{{\tau}}$.
\end{proof}

\

\section{Inferring $\mathbb{F}$ from $\mathbb{G}$}\label{condB}

This last section is devoted to a discussion on the relevance of the progressive enlargement of filtration technique used in credit risk modeling. We consider the question whether the filtration $\mathbb{F}$ can be inferred from the knowledge of the market information $\mathbb{G}$ and from the default time ${\tau}$.

Our discussions make use of probability measures which can be singular with respect to each other. For this matter, we suppose that $\Omega$ is a Polish space and $\mathcal{B}^0$ is its Borel $\sigma$-algebra such that $\mathcal{B}$ is the $\mathbb{Q}$ completion of $\mathcal{B}^0$. We suppose in addition that there exists a filtration $\mathbb{G}^0=(\mathcal{G}^0_t)_{t\in\mathbb{R}_+}$ of Borel $\sigma$-algebras such that $\mathbb{G}$ is the usual augmentation under $\mathbb{Q}$ of $\mathbb{G}^0_+$. We suppose that ${\tau}$ is a $\mathbb{G}^0_+$ stopping time. We consider the following condition. 

\bassumption
\textbf{Condition ($B^0$)} There exists a filtration  $\mathbb{F}^0=(\mathcal{F}^0_s)_{s\in\mathbb{R}_+}$ such that $$
\mathcal{G}^0_{s+}=\cap_{s'>s}(\mathcal{F}^0_{s'}\vee\sigma({\tau}\wedge s')), \ s\in\mathbb{R}_+.
$$ 
\eassumption

\bd
\textbf{Saturation property.} Let $\mathcal{H}$ be a sub-$\sigma$-algebra in $\mathcal{G}^0_\infty$. $\mathcal{H}$ satisfies the saturation property, if, for any probability measure $\mathbb{P}$ on $\mathcal{B}^0$ which makes $\sigma({\tau})$ a trivial $\sigma$-algebra, for any element $A'$ in $\mathcal{G}^0_\infty$, there exists an element $A$ in $\mathcal{H}$ such that $\mathbb{P}[A\Delta A']=0$.

If the above property holds only for a specific probability measure $\mathbb{P}$, we mention it as the saturation property under $\mathbb{P}$.
\ed

We have immediately the following lemma.

\bl
Suppose the condition($B^0$). The $\sigma$-algebra $\mathcal{F}^0_\infty$ satisfies the saturation property. 
\el

We now look at the implications of the saturation property. For doing so, we introduce $\mathfrak{K}(t,d\omega)$ to denote (a version of) the regular conditional probability measure on $\mathcal{B}^0$ given ${\tau}=t$ under $\mathbb{Q}$ (cf. \cite[Chapter 13]{KS} or \cite[Theorem 89.1]{RW}). Let $\nu$ be the law of ${\tau}$ under $\mathbb{Q}$. Note that, for $\nu$-almost all $t$, $\mathfrak{K}(t)[{\tau}\neq t]=0$.

\bl
Let $\mathcal{H}$ be a sub-$\sigma$-algebra in $\mathcal{G}^0_\infty$, which is separable. Suppose the saturation property for $\mathcal{H}$. Then, $$
\mathcal{G}^0_\infty\vee\{\mbox{\small $(\mathcal{G}^0_\infty,\mathbb{Q})$-null sets}\}
=
(\mathcal{H}\vee\sigma({\tau}))\vee\{\mbox{\small $(\mathcal{G}^0_\infty,\mathbb{Q})$-null sets}\}.
$$
\el

\begin{proof}
According to \cite[Lemma 3, Remarque 1]{SY}, for bounded $\mathcal{G}^0_\infty$ measurable function $f$, there exists a $\mathcal{H}\otimes\mathcal{B}[0,\infty]$ measurable map $\Phi(\omega,t)$ which is, for every fixed $t$, a version of $\mathbb{E}_{\mathfrak{K}(t)}[f|\mathcal{H}](\omega)$. We have$$
\mathbb{Q}[f\neq \Phi(\cdot,{\tau})]
=
\int_{[0,\infty]}\nu(dt)
\mathfrak{K}(t)[f\neq \Phi(\cdot,t)]
=0,
$$
because, for $\nu$-almost all $t$, $f$ is $\mathfrak{K}(t)$-almost surely equal to a $\mathcal{H}$ measurable function, thank to the saturation property.
\end{proof}

\bl\label{Hfiltr}
Let $\mathcal{H}$ be a sub-$\sigma$-algebra in $\mathcal{G}^0_\infty$. Let $\mathbb{P}$ be a probability measure on $\mathcal{B}^0$ under which $\sigma({\tau})$ is a trivial $\sigma$-algebra. Suppose the saturation property under $\mathbb{P}$ of $\mathcal{H}$. Let $$
\mathcal{H}_s=\{A\in\mathcal{H}: \exists A'\in\mathcal{G}^0_s, \mathbb{P}[A\Delta A']=0\}, s\in\mathbb{R}_+.
$$
Then, $\mathbb{H}=(\mathcal{H}_s)_{s\in\mathbb{R}_+}$ is a filtration. The $\mathbb{P}$-augmentation of $\mathbb{H}_+$ coincides with that of $\mathbb{G}^0_+$.
\el

The proof of the above lemma is immediate. In the next lemma we consider a particular absolute continuity and its consequence. 

\bd
\textbf{Condition ($B^1$)} Let $\mathcal{H}$ be a sub-$\sigma$-algebra in $\mathcal{G}^0_\infty$ and $\mathbb{P}$ be a probability measure on $\mathcal{B}^0$. We say that $\mathbb{P}$ satisfies condition ($B^1$) on $\mathcal{H}$, if 

\vspace{-13pt}

\ebe
\item[.]
$\sigma({\tau})$ is a trivial $\sigma$-algebra under $\mathbb{P}$ and 
\item[.]
$\mathbb{Q}$ is absolutely continuous with respect to $\mathbb{P}$ on $\mathcal{H}$. 
\dbe

\ed

\bl\label{absctn}
Let $\mathcal{H}$ be a sub-$\sigma$-algebra in $\mathcal{G}^0_\infty$. Suppose that there exists a probability measure $\mathbb{P}$ satisfying condition ($B^1$) on $\mathcal{H}$. Then, ${\tau}$ is not $\mathcal{H}$ measurable, whenever ${\tau}$ is $\mathbb{Q}$ non trivial.
\el

\begin{proof}
Under $\mathbb{P}$, the distribution function of ${\tau}$ is a function of the form $\ind_{[a,\infty)}$ for some $a\in[0,\infty]$, i.e., $\mathbb{P}[{\tau}\neq a]=0$. But $\mathbb{Q}[{\tau}\neq a]>0$ for non trivial ${\tau}$, which means that ${\tau}$ is not $\mathcal{H}$ measurable. 
\end{proof}

In next theorem, we discuss how to detect $\mathbb{F}^0$ if condition ($B^0$) is satisfied with $\mathbb{F}^0$.

\bethe\label{findF}
Suppose condition($B^0$) with a filtration $\mathbb{F}^0$. Suppose that there exist $\mathcal{H}$ a sub-$\sigma$-algebra in $\mathcal{G}^0_\infty$ and $\mathbb{P}$ a probability measure satisfying condition ($B^1$) on $\mathcal{H}$, such that $\mathcal{F}^0_{\infty}\subset\mathcal{H}$. Then, for the filtration $(\mathcal{H}_s)_{s\in\mathbb{R}_+}$ defined in Lemma \ref{Hfiltr}, we have
$$
\dcb
\mathcal{F}^0_\infty\vee\{\mbox{\small $(\mathcal{H},\mathbb{Q})$-null sets}\}=\mathcal{H}\vee\{\mbox{\small $(\mathcal{H},\mathbb{Q})$-null sets}\},\\
\mathcal{F}^0_{s+}\vee\{\mbox{\small $(\mathcal{H},\mathbb{Q})$-null sets}\}=\mathcal{H}_{s+}\vee\{\mbox{\small $(\mathcal{H},\mathbb{Q})$-null sets}\}, s\in\mathbb{R}_+.
\dce
$$
\ethe

\begin{proof}
The left hand side $\sigma$-algebra of the first identity is clearly contained in the right hand side $\sigma$-algebra. In the opposite direction, as $\mathcal{H}\subset\mathcal{G}^0_\infty$, any element in $\mathcal{H}$ is $\mathbb{P}$-almost surely equal to an element in $\mathcal{F}^0_\infty$, because $\sigma({\tau})$ is trivial under $\mathbb{P}$. This equality holds also $(\mathcal{H},\mathbb{Q})$-almost surely, proving the first identity.

Consider the second identity. Let $s<s'$ be positive real numbers. Every element in $\mathcal{F}^0_{s}$ ($\subset \mathcal{G}^0_{s+}\subset \mathcal{G}^0_{s'}$) differs from an element of $\mathcal{H}_{s'}$ only by a $\mathbb{P}$ null set. As $\mathcal{F}^0_{s}$ is contained in $\mathcal{H}$, this difference is $(\mathcal{H},\mathbb{Q})$ negligible. Hence
$$
\mathcal{F}^0_{s+}\vee\{\mbox{\small $(\mathcal{H},\mathbb{Q})$-null sets}\}
\subset
\mathcal{H}_{s+}\vee\{\mbox{\small $(\mathcal{H},\mathbb{Q})$-null sets}\}.
$$
Conversely, an element in $\mathcal{H}_{s}$ differs from an element in $\mathcal{G}^0_{s}$ only by a $\mathbb{P}$ null set, and, as $\sigma({\tau})$ is trivial under $\mathbb{P}$, differs from an element of $\mathcal{F}^0_{s'}$ only by a $\mathbb{P}$ null set, which is also a $(\mathcal{H},\mathbb{Q})$ null set. We prove
$$
\mathcal{F}^0_{s+}\vee\{\mbox{\small $(\mathcal{H},\mathbb{Q})$-null sets}\}
\supset
\mathcal{H}_{s+}\vee\{\mbox{\small $(\mathcal{H},\mathbb{Q})$-null sets}\}.
$$
\end{proof}

\brem
When a condition ($B^0$) is assumed without the knowledge of $\mathbb{F}^0$, Theorem \ref{findF} gives a method to recover the filtration $\mathbb{F}^0$, whenever $\mathcal{F}^0_\infty$ can be located and a probability measure $\mathbb{P}$ can be found to satisfy condition ($B^1$) on $\mathcal{F}^0_\infty$. Another interpretation of Theorem \ref{findF} is that, each time one has a saturating $\mathcal{H}$ and a $\mathbb{P}$ satisfying condition ($B^1$) on $\mathcal{H}$, one should check if the filtration $\mathbb{H}$ defined in Lemma \ref{Hfiltr} satisfies condition ($B^0$).
\erem

We give below two examples of condition ($B^0$) where, for some $t$, $\mathfrak{K}(t)$ satisfies condition ($B^1$) on $\mathcal{F}^0_\infty$. The first example is based on the density hypothesis (cf. \cite{ejj, pham}).

\bassumption
\textbf{Strict density hypothesis.}
The condition($B^0$) holds. There exists a probability measure $\nu$ on $\mathcal{B}[0,\infty]$ and a strictly positive $\mathcal{B}[0,\infty]\otimes\mathcal{F}^0_\infty$ measurable function $p(t,\omega)$, $(t,\omega)\in[0,\infty]\times\Omega$, such that$$
\mathbb{E}_\mathbb{Q}[h({\tau})|\mathcal{F}^0_\infty](\omega)=\int_{[0,\infty]}h(t)p(t,\omega)\nu(dt)
$$
for any bounded Borel function $h$. 
\eassumption

If the strict density hypothesis is satisfied, the probability measure $\mathbb{P}=\frac{1}{p({\tau},\cdot)}{_\centerdot}\mathbb{Q}$ makes ${\tau}$ independent of $\mathcal{F}^0_\infty$, with which we compute
$$
\mathbb{E}_\mathbb{Q}[f|\sigma({\tau})]
=
\frac{\mathbb{E}_{\mathbb{P}}[f p({\tau},\cdot)|\sigma({\tau})]}{\mathbb{E}_{\mathbb{P}}[ p({\tau},\cdot)|\sigma({\tau})]}
=
\left.\frac{\mathbb{E}_{\mathbb{Q}}[f p(t,\cdot)]}{\mathbb{E}_{\mathbb{Q}}[ p(t,\cdot)]}\right|_{t={\tau}}
$$
for any non negative $\mathcal{F}^0_\infty$ measurable function $f$. This computation means that, for $\nu$ almost all $t$, $\left.\frac{\mathbb{E}_{\mathbb{Q}}[f p(t,\cdot)]}{\mathbb{E}_{\mathbb{Q}}[ p(t,\cdot)]}\right|_{t={\tau}}$ coincides with $\mathfrak{K}(t)$ restricted on $\mathcal{F}^0_\infty$, i.e., this restriction of $\mathfrak{K}(t)$ is a measure equivalent to $\mathbb{Q}$ with a density function $\frac{p(t,\cdot)}{\mathbb{E}_{\mathbb{Q}}[ p(t,\cdot)]}$.

The next example is about Cox time (cf. \cite{BJR}).

\bassumption
\textbf{Cox time.} Suppose the condition($B^0$). Suppose that there exists a exponential random time $\xi$ independent of $\mathcal{F}^0_\infty$ and a c\`adl\`ag non decreasing $\mathbb{F}^0_+$ adapted process $(a_s)_{s\in\mathbb{R}_+}$ such that $a_0=0$ and $$
{\tau}=\inf\{s>0: a_s>\xi\}.
$$
\eassumption

Consider a Cox time. For simplicity, suppose that $a_\infty=\infty$. We have $$
\mathbb{E}[{\tau}\geq s|\mathcal{F}^0_\infty]
=
\mathbb{E}[a_{s-}\leq \xi |\mathcal{F}^0_\infty]
=
e^{-a_{s-}}.
$$
and therefore $\mathbb{E}[{\tau}> s|\mathcal{F}^0_\infty]=e^{-a_{s}}$. Consequently, if $da_s$ is absolutely continuous with respect to the Lebesgue's measure with a strictly positive density function, we are back to the situation of the strict density hypothesis. We now construct an example where the strict density hypothesis is not satisfied.

Let $T$ be an $\mathbb{F}$ stopping time whose law $\mu^T$ is diffuse and singular to the Lebesgue measure and let $a_s=s+\ind_{\{T\leq s\}}, s\in\mathbb{R}_+$. Then, by the above analysis, $$
\mathbb{E}[{\tau}> s|\mathcal{F}^0_\infty]
=e^{-(s+\ind_{\{T\leq s\}})}
=
\int_s^T e^{-t}dt + (e^{-T}-e^{-1-T})\ind_{\{s\leq T\}}+\int_{s\vee T}^\infty e^{-1-t}dt. 
$$
Let $A\subset[0,\infty]$ be a Borel set and let $f$ be a non negative $\mathcal{F}^0_\infty$ measurable function. Denote by $\mathbb{E}_{T=t}$ the conditional expectation on $\mathcal{F}^0_\infty$ given $T=t$. Let $B$ be a support of $\mu^T$ having null Lebesgue measure. We compute$$
\dcb
\mathbb{E}[f\ind_{\{{\tau}\in A\}}]
&=&
\mathbb{E}[f\ind_{\{{\tau}\in A\}}\ind_{\{{\tau}<T\}}]
+\mathbb{E}[f\ind_{\{{\tau}\in A\}}\ind_{\{{\tau}=T\}}]
+\mathbb{E}[f\ind_{\{{\tau}\in A\}}\ind_{\{{\tau}>T\}}]\\

&=&
\mathbb{E}[f\int_0^T\ind_{\{t\in A\}}e^{-t}dt]
+\mathbb{E}[f\ind_{\{T\in A\}}(e^{-T}-e^{-1-T})]
+\mathbb{E}[f\int_T^\infty\ind_{\{t\in A\}}e^{-1-t}dt]\\

&=&
\int_0^\infty\ind_{\{t\in A\}}e^{-t}dt\mathbb{E}[(\ind_{\{t\leq T\}}+e^{-1}\ind_{\{t> T\}})f]\\
&&+\int_0^\infty\ind_{\{t\in A\}}(e^{-t}-e^{-1-t})\mu^T(dt)\mathbb{E}_{T=t}[f]\\

&=&
\int_0^\infty\ind_{\{t\in A\}}\left(e^{-t}\mathbb{E}[(\ind_{\{t\leq T\}}+e^{-1}\ind_{\{t> T\}})]\ind_{\{t\in B^c\}}
+(e^{-t}-e^{-1-t})\ind_{\{t\in B\}}\right)(dt+\mu^T(dt))\\

&&\left(\ind_{\{t\in B^c\}}\frac{\mathbb{E}[(\ind_{\{t\leq T\}}+e^{-1}\ind_{\{t> T\}})f]}{\mathbb{E}[(\ind_{\{t\leq T\}}+e^{-1}\ind_{\{t> T\}})]}+\ind_{\{t\in B\}}\mathbb{E}_{T=t}[f]\right).
\dce
$$
This computation shows that the law of ${\tau}$ is determined by the expression
$$
\int_0^\infty\ind_{\{t\in A\}}\left(e^{-t}\mathbb{E}[(\ind_{\{t\leq T\}}+e^{-1}\ind_{\{t> T\}})]\ind_{\{t\in B^c\}}
+(e^{-t}-e^{-1-t})\ind_{\{t\in B\}}\right)(dt+\mu^T(dt))
$$
and the conditional expectation on $\mathcal{F}^0_\infty$ given ${\tau}=t$ is provided by the expression
$$
\mathfrak{K}(t)[f]
=
\left(\ind_{\{t\in B^c\}}\frac{\mathbb{E}[(\ind_{\{t\leq T\}}+e^{-1}\ind_{\{t> T\}})f]}{\mathbb{E}[(\ind_{\{t\leq T\}}+e^{-1}\ind_{\{t> T\}})]}+\ind_{\{t\in B\}}\mathbb{E}_{T=t}[f]\right).
$$
We see then that there exists $t\in B^c$ such that $\mathbb{Q}\ll \mathfrak{K}(t)$ on $\mathcal{F}^0_\infty$. 

\

{}

\end{document}